\documentclass[aps,floatfix,showpacs,preprintnumbers,amsmath,amssymb]{revtex4}
\usepackage{natbib}
\usepackage{dcolumn}
\usepackage{graphicx}
\usepackage{enumerate}
\usepackage{inputenc}
\usepackage{fontenc}
\usepackage{float}
\usepackage{color}
\usepackage{ulem}
\usepackage{hyperref}

\newcommand{\mnras}{Mon.~Not.~R.~Astron.~Soc.~}

\newcommand{\phrd}{Phys.~Rev.~D.~}
\newcommand{\jcap}{J.~Cosmol.~Astropart.~Phys.~}
\newcommand{\be}{\begin{equation}}
\newcommand{\ee}{\end{equation}}
\newcommand{\bea}{\begin{eqnarray}}
\newcommand{\eea}{\end{eqnarray}}
\providecommand{\abs}[1]{\lvert#1\rvert}
\def\[{\begin{equation}}
\def\]{\end{equation}}
\begin{document}
\title{Effects of Dark Energy Perturbations on Cosmological Tests of General Relativity}
\author{Jason N.  Dossett$^{1,2}$\footnote{Electronic address: j.dossett@uq.edu.au}, Mustapha Ishak$^1$\footnote{Electronic address: mishak@utdallas.edu}}
\affiliation{
$^1$Department of Physics, The University of Texas at Dallas, Richardson, TX 75083, USA\\
$^2$School of Mathematics and Physics, University of Queensland, Brisbane, QLD 4072, Australia}
\date{\today}
\begin{abstract}
Cosmological tests to distinguish between dark energy (DE) and modifications to gravity are a promising route to obtain clues on the origin of cosmic acceleration. We study here the robustness of these tests to the presence of DE density, velocity, and anisotropic stress perturbations. We find that the dispersion in the growth index parameter remains small enough to distinguish between extreme cases of DE models and some commonly used modified gravity models. The sign of the slope parameter for a redshift dependent growth index was found to be inconsistent as an additional test in extreme cases of DE models with perturbations. Next, we studied the effect of DE perturbations on the modified growth (MG) parameters that enter the perturbed Einstein equations. We find that while the dark energy perturbations affect the MG parameters, the deviations remain smaller than those due to modified gravity models. Additionally, the deviations due to DE perturbations with a non-zero effective sound speed occur at scale ranges that are completely different than those due to some modified gravity models such as the $f(R)$ models. In the case of modified gravity models with zero anisotropic stress at late times, the simultaneous determination of the effective dark energy equation of state and the MG parameters can provide the distinction between these models and DE. The growth index test was found to be the most robust to these perturbations. The scale dependence of the MG parameters in some cases of modified gravity constitute a clear-cut discriminant regardless of any DE perturbations. In summary, we find that the currently proposed cosmological tests to distinguish between DE and modified gravity are robust to DE perturbations even for extreme cases. This is certainly the case even for DE models with equations of state of DE that fall well outside of current cosmological constraints.
\end{abstract}
\maketitle
\section{Introduction}
The quest to understand the origin of cosmic acceleration has led the scientific community to develop 
methods and tests that can provide clues from observations. First, comes the constraints from cosmological data sets that one can obtain on the dark energy equation of state parameters. For example, depending on whether the dark energy equation of state is $-1$ or not, up to some level of significance, one could infer whether or not the data is consistent with a cosmological constant. A second promising approach that could lead us to learn more about the possible cause of cosmic acceleration is to test whether the acceleration is a sign of some gravity physics that extends or replaces general relativity at cosmological scales. 

Indeed, tests that can help one to distinguish whether cosmic acceleration is due to some dark energy in the universe or, rather, some new gravity physics at cosmological scales have attracted a lot of attention in recent papers. While frameworks and parameters have been discussed in for example \cite{LSS04DEP, Song2005aDEP, KST06DEP, Ishak2006DEP, Linder2005DEP, KoyamaDEP,Zhang2006DEP, ZhangEtal2007DEP, HS07DEP, KS08DEP,GMPDEP, HL07DEP, LC07DEP, GabadadzeDEP, PolarskiDEP,MotohashiDEP, BZ08DEP,JainDEP, WeiDEP,MHH1DEP, Dent2009DEP, GIW09DEP, Ishak2009DEP,Gong2008DEP, Simpson09DEP, Linder11DEP, WuDEP,BakerDEP}, comparison to current and simulated data can be found in a number of references \cite{CCM07DEP, AcquavivaDEP,ThomasDEP,  Daniel2008DEP,Daniel2009DEP, DossettDEP,Daniel2010DEP, MHH2DEP,Bean2010DEP, DL2010DEP, SerraDEP, GongBo2010DEP,LombriserDEP, TorenoDEP, Dossett2011DEP, SongZhao2011DEP, MGCAMB2DEP, ISiTGRDEP, HojjatiPCADEP, Laszlo2011DEP, DIDEP, Simpson2012DEP, Rapetti2012DEP,GMSM09}.
(these are partial list only, see further references therein).

The two most popular ways to distinguish between dark energy and modifications to gravity physics on cosmological scales both make use of parameters that take-on known values in general relativity.  The first uses the growth index, $\gamma$, which characterizes the logarithmic growth rate, $f=d\ln \delta/d\ln a$.  The growth index formalism was first introduced by \cite{PeeblesDEP}, where it was proposed that $f$ could be approximated using the ansatz  $f =\Omega_m^\gamma$. This has been reused and extended in the framework of dark energy by \cite{wangDEP} and was proposed as a way to distinguish between dark energy and modified gravity models by \cite{Linder2005DEP}.  By now, it is well known that the $\Lambda$CDM value of growth index is $\gamma = 6/11$.  Since its introduction as a way to distinguish between dark energy and modifications to gravity there has been a sizable amount of study on using $\gamma$ for this task (see for example: \cite{MotohashiDEP, WuDEP, Gong2008DEP,MHH1DEP, HL07DEP, LC07DEP, PolarskiDEP, Dent2009DEP, GIW09DEP,GMPDEP, Ishak2009DEP, DossettDEP,MHH2DEP, Simpson09DEP, Rapetti2012DEP}).

The second, popular method used to test deviations from general relativity (or the presence of modified gravity) focuses on using parameters that parameterize deviations from known growth equations, primarily the Poisson equation and anisotropy equation.  These parameterizations are usually set up in a way where the modified growth (MG) parameters will take a value of either 1 or 0 in general relativity (see, for example, \cite{ZhangEtal2007DEP,BZ08DEP,JainDEP, SerraDEP,ThomasDEP,CCM07DEP,Daniel2008DEP,Daniel2009DEP,GongBo2010DEP, Bean2010DEP, Daniel2010DEP, DL2010DEP, LombriserDEP, TorenoDEP,Linder11DEP, SongZhao2011DEP, Dossett2011DEP, MGCAMB2DEP, ISiTGRDEP, Laszlo2011DEP, BakerDEP, HojjatiPCADEP, DIDEP, Simpson2012DEP,GMSM09}). 

It was shown in these previous studies that the two tests above can be successful in distinguishing   
dark energy from modified gravity models. However, the question of how dark energy models with 
density and anisotropic stress perturbations could affect the conclusiveness of these tests still needs 
a thorough exploration and is the subject of this paper.  

In section \ref{sec:pert} we briefly review the growth equations for perturbations of the Friedmann-Lema\^{i}tre-Robertson-Walker (FLRW) metric as well as a description of the modified growth (MG) parameters used to detect deviations from general relativity by directly modifying the perturbed Einstein equations.  Then in section \ref{sec:depert} we give a quick description of the equations typically used to describe dark energy perturbations (We do not consider dark energy models that have exotic interactions with other matter species).  We derive new expressions that relate the dark energy perturbations to the MG parameters in section \ref{sec:MGderiv}.  Next, in section \ref{sec:Res}, we explore the influence of the dark energy perturbations on the various parameters used in tests to distinguish between general relativity + dark energy and modifications to gravity, particularly, the growth index and the MG parameters $Q$ and $R$.  First, in section \ref{sec:pertonly} we look at the influence dark energy models with only density and velocity perturbations. Then in section \ref{sec:pertpi} we look at dark energy models that additionally include anisotropic stress perturbations. We complete our analysis in section \ref{sec:cs2} by exploring the impact changes to the effective sound speed of dark energy perturbations have on the various tests. Finally, in section \ref{sec:conclude} we summarize our results and make some concluding remarks.

%
%
\section{Background}
We briefly describe here the formalism that we use in this paper (see, for example, \cite{MaDEP,ZSBDEP,ISiTGRDEP} for more detailed descriptions). It is perhaps useful to mention here that we take the point of view that cosmic acceleration is not synonym of dark energy but rather cosmic acceleration can be caused by: 

1) dark energy that invoke some extra component in the makeup of the total energy density of the universe. With this component accounted for in the energy momentum tensor, Einstein's Equations of General Relativity are then used to describe gravity in the universe leading to an accelerated expansion. Or 

2) modified gravity models that explain cosmic acceleration by changing the gravity theory of the universe, rather than invoking some unknown energy content in the energy momentum tensor.  If a modified gravity model were the correct gravity model, it would mean that General Relativity does not adequately describe gravitational interactions at cosmological scales. In other words the coupling between dark + baryonic matter and spacetime curvature is changed via some field equations beyond Einstein's General Relativity. 

The distinction between category (1) and  category (2) above is the point of view that we take in our paper while other points view where these are the same have also been taken. It is true that some modified gravity models can be subject to a transformation to GR plus a scalar field but not all modified gravity models can be transformed in such a way. And if so, this does not provide exactly the same physical explanation as that the universe is filled with some dark energy component. One could perhaps argue the opposite but we do not take that point of view in this paper. 
\subsection{Growth of metric perturbations\label{sec:pert}}

The flat, perturbed FLRW metric written in the conformal Newtonian gauge is given by
\be
ds^2=a(\tau)^2[-(1+2\psi)d\tau^2+(1-2\phi)dx^idx_j],
\label{eq:FLRWNewt}
\ee
where $\phi$ and $\psi$ are scalar potentials describing the scalar mode of the metric perturbations, $\tau$ is conformal time, $a(\tau)$ is the scale factor normalized to one today, and the $x_i$'s are the comoving coordinates. 

Using Einstein's field equations we can quickly obtain the Poisson equation and the anisotropy equation, respectively:
\bea
k^2\phi  &=&-4\pi G a^2\sum_i \rho_i \Delta_i,
\label{eq:Poisson}\\
k^2(\psi-\phi) &=& -8 \pi G a^2\sum_i \rho_i w_i\Pi_i.
\label{eq:2ndEin}
\eea
In these equations, $i$ denotes an individual matter species, $\rho_i$ is the density, $\Delta_i$ is the gauge-invariant, rest-frame overdensity, $\Pi_i$ is the anisotropic stress perturbation, and $w=P/\rho$ is the equation of state of the fluid. Here we have chosen to use the anisotropic stress perturbation, $\Pi$, which is related to the shear stress, $\sigma$, by $\sigma_\alpha=\frac{2}{3}\Pi_\alpha w_\alpha /(1 + w_\alpha)$ \cite{MaDEP}. 

For our modified growth equations we will use the formalism introduced by \cite{Bean2010DEP}.  These equations read
\bea
k^2\phi  &=& -4\pi G a^2\sum_i \rho_i \Delta_i \,  Q
\label{eq:PoissonMod}\\
k^2(\psi-R\,\phi) &=& -8 \pi G  a^2\sum_i \rho_i w_i \Pi_i \, Q,
\label{eq:Mod2ndEin}
\eea
where $Q$ and $R$ are the modified growth parameters (MG parameters).  We write separately equations (\ref{eq:Poisson})-(\ref{eq:2ndEin}) and (\ref{eq:PoissonMod})-(\ref{eq:Mod2ndEin}) in order to avoid any ambiguity when we refer extensively to each set in the following sections. A modification to the Poisson equation is quantified by the parameter $Q$, while the gravitational slip (a term coined by \cite{CCM07DEP}) is quantified by the parameter $R$ (at late times, assuming anisotropic stress is negligible, $\psi= R\phi$).
\subsection{Dark energy perturbations \label{sec:depert}}
\subsubsection{Dark energy models with density perturbations}
Before discussing our results, it is necessary to discuss the standard evolution equations for dark energy perturbations.  As in our description of the modified growth equations we will be working in the conformal Newtonian gauge. Enforcing conservation of energy momentum on a perturbed fluid gives the following two equations \cite{MaDEP}:
\be
\dot{\delta}  =  -(1+w)(\theta-3\dot{\phi})+3\mathcal{H}(w-\frac{\delta P}{\delta\rho})\delta
\label{eq:deltaevo}
\ee
\vspace{-\baselineskip}
\be
\dot{\theta}= -\mathcal{H}(1-3w)\theta-\frac{\dot{w}}{1+w}\theta +\frac{\delta P/\delta\rho}{1+w}k^2\delta+k^2\psi,
\label{eq:thataevo}
\ee
where $\delta$ is the fractional overdensity, $\delta\rho/\rho$, $\theta$ is the divergence of the peculiar velocity, and $P$ is the pressure.  

To handle perturbations of dark energy, it is useful to define an effective sound speed of dark energy perturbations, $c_s$, such that \cite{dPHLDEP,BDDEP,Hu98DEP}:
\be
\frac{\delta P}{\delta \rho }\delta \equiv \frac{\delta P}{\rho} = c_s^2\delta+3\mathcal{H}(1+w)(c_s^2-c_a^2)\frac{\theta}{k^2},
\label{eq:delPdelrho}
\ee
where $c_a$ is the adiabatic sound speed, given by
\be
c_a^2 = \frac{\dot{P}}{\dot{\rho}} = w - \frac{\dot{w}}{3\mathcal{H}(1+w)}.
\label{eq:ca2}
\ee

Now subbing Eq. (\ref{eq:delPdelrho}) into Eqs. (\ref{eq:deltaevo}) and (\ref{eq:thataevo}) we have for the evolution equations for dark energy perturbations with an effective sound speed, $c_s$ \cite{dPHLDEP}:

\be
\dot{\delta} = -(1+w)\Big\{\left[k^2+9\mathcal{H}^2(c_s^2-c_a^2)\right]\frac{\theta}{k^2}-3\dot{\phi}\Big\}+3\mathcal{H}(w-c_s^2)\delta
\label{eq:DEdeltadot}
\ee
\vspace{-\baselineskip}
\be
\dot{\theta}= (3c_s^2-1)\mathcal{H}\theta +k^2\frac{c_s^2\delta }{1+w}+k^2\psi.
\label{eq:DEthetadotnopi}
\ee
\subsubsection{Dark energy models with density and anisotropic stress perturbations\label{sec:DEpi}}
Above we have considered only dark energy models where the dark energy was modeled as a perfect fluid.  In the most general case, one should also consider dark energy models with an anisotropic stress, $\Pi$.  Such models have been discussed previously in, for example, \cite{Hu98DEP,KM05DEP,MKKG07DEP}.  For brevity, here we will quickly review the relevant equations for these models, however a more in depth discussion of these models is available in the aforementioned references.     

First, we should define the evolution equation for the anisotropic stress perturbation.  This was first given in \cite{Hu98DEP}. In the conformal Newtonian gauge this equation is written:
\be
\dot{\Pi} + 3\mathcal{H}\Pi=4\frac{c_{\rm vis}^2}{w}\theta, 
\label{eq:pidot}
\ee
where $c_{\rm vis}^2$ is the viscosity parameter. As discussed in \cite{KM05DEP}, in order to produce stable solutions, $c_{\rm vis}^2$ must have the same sign as $(1+w)$.  

Next we must consider the effect that the anisotropic stress perturbations in the dark energy will have on the evolution of the other dark energy perturbation variables.  In \cite{Hu98DEP} it is shown that $\delta$ is only indirectly affected, while the evolution of $\theta$ is directly modified and given by 
\be
\dot{\theta}= (3c_s^2-1)\mathcal{H}\theta +k^2\frac{c_s^2\delta }{1+w}+k^2\left(\psi-\frac{2}{3}\frac{w}{1+w}\Pi\right).
\label{eq:DEthetadot}
\ee

\section{Relations between dark energy perturbations and the MG parameters\label{sec:MGderiv}}

We will now explore how the effects dark energy perturbations can mimic any possible departures of MG parameters introduced in Eqs. (\ref{eq:PoissonMod}) and (\ref{eq:Mod2ndEin}) from their value of unity in $\Lambda$CDM.  While the relationship between the dark energy perturbations and the growth index, $\gamma$, is hard to explore analytically, with the MG parameters we can actually derive analytic expressions in terms of already defined variables.  

Before deriving these expressions, let us first discuss how tests with these parameters are performed.  When performing tests using MG parameters the usual approach is to make the assumption that we are in the presence of a $\Lambda$CDM model and look for deviations from that model using MG parameters such as those in Eqs. (\ref{eq:PoissonMod}) and (\ref{eq:Mod2ndEin}). Given that we are assuming a $\Lambda$CDM background, when we look at these equations, none of the quantities on the right-hand side (RHS) of Eqs. (\ref{eq:PoissonMod}) and (\ref{eq:Mod2ndEin}) are dark energy quantities since dark energy does not have perturbations or anisotropic stress perturbations in $\Lambda$CDM. 

Now to see how dark energy perturbations affect the MG parameters, let us now consider a case where the true underlying background model does allow for dark energy to have perturbations and shear.  In this case, the underlying model has potentials governed by Eqs. (\ref{eq:Poisson}) and (\ref{eq:2ndEin}), where the quantities on the RHS do include dark energy quantities. We will denote these dark energy quantities with a subscript $_{DE}$ below.  We can calculate how the effects of these dark energy perturbations can mimic the presence of the parameters $Q$ and $R$ by noticing that when performing these tests the metric potentials from the modified growth equations must match the metric potentials of the true underlying model.  

Since the left-hand sides of Eqs. (\ref{eq:Poisson}) and (\ref{eq:PoissonMod}) are the same, we can simply set the right-hand sides of these equations equal to one another and solve for $Q$. Separating out the dark energy perturbations in Eq. (\ref{eq:Poisson}), we have:
\bea
-Q\,4\pi G a^2\sum_{i\ne DE} \rho_i \Delta_i\,  &=& -4\pi G a^2\sum_{i\ne DE} \rho_i \Delta_i - 4\pi G a^2 \rho_{_{DE}} \Delta_{DE}\\
\Rightarrow Q &=& 1+\frac{\rho_{_{DE}} \Delta_{DE}}{\sum\limits_{i\ne DE} \rho_i \Delta_i}.
\label{eq:Qeval}
\eea

Obtaining an expression for $R$ is a little more cumbersome as we must first evaluate an expressions for $\psi$. By combining Eq. (\ref{eq:Poisson}) with Eq. (\ref{eq:2ndEin}) and Eq. (\ref{eq:PoissonMod}) with Eq. (\ref{eq:Mod2ndEin}) we obtain:
\be
k^2\psi = -\sum_{i\ne DE}\tilde{\rho}_i\Big[w_i\Pi_i +\frac{\Delta_i}{2}\Big] -\tilde{\rho}_{_{DE}} \Big[w_{_{DE}}\Pi_{_{DE}}+\frac{\Delta_{DE}}{2}\Big], \label{eq:psiNorm}
\ee
\vspace{-\baselineskip}
\be
k^2\psi = -Q\sum_{i\ne DE}\tilde{\rho}_i\Big[w_i\Pi_i +R\frac{\Delta_i}{2}\Big],
\label{eq:psiMod}
\ee
where $\tilde{\rho}_\alpha = 8\pi G a^2\rho_\alpha$.  Now equating the RHS of these equations, subbing in for $Q$ using Eq. (\ref{eq:Qeval}), and solving for $R$ gives
\be
R =1+  2\frac{\rho_{_{DE}}w_{_{DE}}\Pi_{_{DE}}-\frac{\rho_{_{DE}}\Delta_{DE}}{\sum\limits_{i\ne DE}\rho_i\Delta_i}\sum\limits_{i\ne DE}\rho_i w_i\Pi_i}{\sum\limits_{i\ne DE}\rho_i\Delta_i+\rho_{_{DE}}\Delta_{DE}}.
\label{eq:Reval}
\ee
This equation shows that, at late times with the assumption that ordinary matter has negligible anisotropic stress, the only way we could see an effect on $R$ is if dark energy has some type of anisotropic stress. 

Taking Eqs. (\ref{eq:Qeval}) and (\ref{eq:Reval}) together, we can see that neglecting a dark energy model with perturbations and anisotropic stress could, in fact, influence our constraints on the MG parameters $Q$ and $R$. We will explore the magnitude of these effects.

\section{Analysis and Results \label{sec:Res}}
For the numerical part of the analysis, we use a modified version of the January 2012 release of the publicly available code \texttt{CAMB} \cite{LewisCAMBDEP}.  We modify \texttt{CAMB} so that we can introduce a varying dark energy equation of state according the parameterization $w(a) = w_0 + w_a(1-a)$  \cite{CPLCPDEP,CPLLDEP}.   Such a parameterization of the equation of state does not allow for rapidly oscillating equations of state such as those that would be observed for models seen in \cite{AZBDEP}.  We then modify the portion of this codes that evolves the dark energy perturbations to allow for such an equation of state.  For the portion of our work where we look at the effect DE perturbations have on MG parameters, we leave the evolution of the DE perturbations to be governed by Eqs. (\ref{eq:DEdeltadot}) and (\ref{eq:DEthetadot}).

We consider a wide range of dark energy models.  For a constant equation of state, we consider models from $w = -1.35$ to $w = -0.65$, incrementing $w$ in steps of $0.05$.  These limits are in excess of the $4\sigma$ limits from the latest WMAP9 cosmological constraints \cite{WMAP9DEP}. For models with a variable equation of state, we avoid models that cross the phantom divide and choose 6 models in total. We write the equation of state for dark energy model using the convention $(w_0,w_a)$.  First, three models that are just within the current WMAP9 $95\%$ confidence limits: a quintessence model $(-0.95,0.10)$ and two phantom models, $(-1.30,0.20)$ and $(-1.10,-0.40)$.  We also consider two phantom models with parameters within the $68\%$ confidence limits: $(-1.05,-0.15)$ and $(-1.20,0.15)$.  Finally, we include a model with an equation of state that corresponds to supergravity (SUGRA) \cite{BMDEP}, $(-0.80,0.30)$ \cite{WADEP}.  This last model has a $w_0$ that is excluded by $3\sigma$ using the combined WMAP9 results. To increase readability we do not include every single one of these models in all of our plots, rather, only the limiting and intermediate cases.

\subsection{Effects of dark energy models with only density and velocity perturbations\label{sec:pertonly}}
We will first look at the effect that dark energy models with only density and velocity perturbations have on the various tests of gravity.  As such, for each of the dark energy models considered, we evolve the perturbations according to Eqs. (\ref{eq:DEdeltadot}) and (\ref{eq:DEthetadotnopi}).

\begin{figure}[t!]
\centering
\begin{tabular}{|c|}
\hline 
{\includegraphics[width=3.5in,height=1.95in,angle=0]{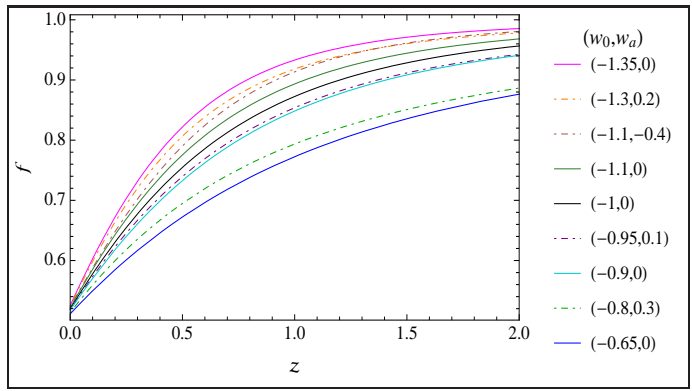}}\\ \hline
\end{tabular}
\caption{\label{f:f} 
We plot the logarithmic growth rate $f$ for various dark energy models where dark energy perturbations are allowed.  The dark energy equation of state for the various models evolves as   $w = w_0 + w_a(1-a)$.  For reference the logarithmic growth rate, $f$, is shown for a $\Lambda$CDM model as a solid black line.  The legend lists the various models used and is ordered according to their values at $z=1$.  While $f$ shows a bit of dispersion for the various models shown here, we see in Fig. \ref{f:gamma} that the growth index parameters show very little dispersion for all of the models shown.} 

\end{figure}

\subsubsection{Impact on the growth index \label{sec:POgi}}
We first start by exploring the effect that the dark energy perturbations considered have on the growth index parameter, $\gamma$.  In \texttt{CAMB}, at each evaluation step, we output: the mass averaged fractional matter overdensity, $\delta_m = (\rho_c \delta_c+\rho_b \delta_b)/(\rho_c+\rho_c)$; $\Omega_m(a) = \Omega_m a^{-3}/(H/H_0)^2$; the wave-number, $k$; and the scale factor, $a$.  We then input this table of values into Mathematica\textsuperscript{\textregistered} and build an interpolating function for $\delta_m(a)$ and $\Omega_m(a)$.  Using this interpolating function, we can quickly arrive at $f$ by using:
\be
f(a) = \frac{d\ln \delta}{d \ln a} = \frac{a\, d \delta}{\delta\,da}
\ee

We evaluate $f$ at $k=0.02$, which is the $k$ at which the amplitude of primordial curvature perturbations, $\mathcal{R}$, was normalized in \cite{WMAP5DEP} (in \cite{WMAP74DEP} $\mathcal{R}$ was normalized at $k =0.027$). We checked and found that the value of $f$ does not change significantly for larger values of $k$. In Fig. \ref{f:f}, we plot the logarithmic growth rate as a function of redshift, $z = 1/a -1$, which we obtain from this interpolation method. 

We fit two parameterizations of $\gamma$ to the ansatz $f(z) = \Omega_m(z)^{\gamma}$.  First, we fit the standard $\gamma = constant$ and then we fit a redshift dependent parameterization. Redshift dependent parameterizations of $\gamma$ such as those introduced by \cite{PolarskiDEP,WuDEP} have been shown to more accurately reproduce the true behavior of $f$ than the constant form.  One example of these parameterization reads
\be
\gamma(z)\, =\, \gamma_0\,+\, \gamma' z, 
\label{gamma1}
\ee
where $\gamma_0$ is the value of the growth index today, and $\gamma'\equiv \frac{d\gamma}{dz}(z=0)$.

Here we choose to use the parameterization first introduced in \cite{DossettDEP}.  This exponential parameterization is written as 
\be
\gamma(z)\, =\, \gamma_e\,+\, \gamma_b\, e^{-z/0.61},
\label{eq:gammaDEP}
\ee
and provides similar results to the parameterization (\ref{gamma1}) above, but it picks out more accurately the high redshift values of the growth index parameter, $\gamma_e$, while still having a slope parameter, $\gamma_b$, that can be used to distinguish between different models of gravity.   

\begin{figure}[t!]
\centering
\begin{tabular}{|c|}
\hline 
{\includegraphics[width=3.45in,height=1.92in,angle=0]{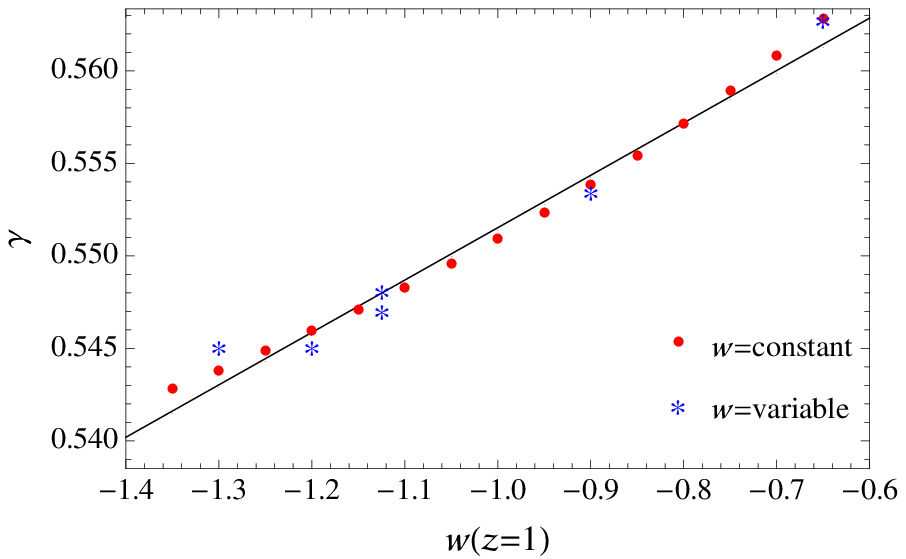}}\\ \hline
\end{tabular}
\\
\begin{tabular}{|c|c|}
\hline
{\includegraphics[width=3.45in,height=1.92in,angle=0]{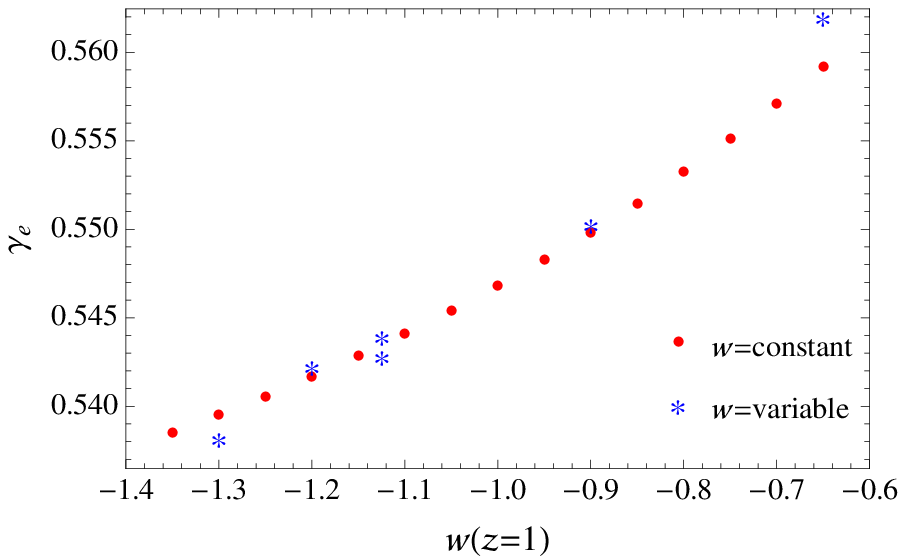}} &
{\includegraphics[width=3.45in,height=1.92in,angle=0]{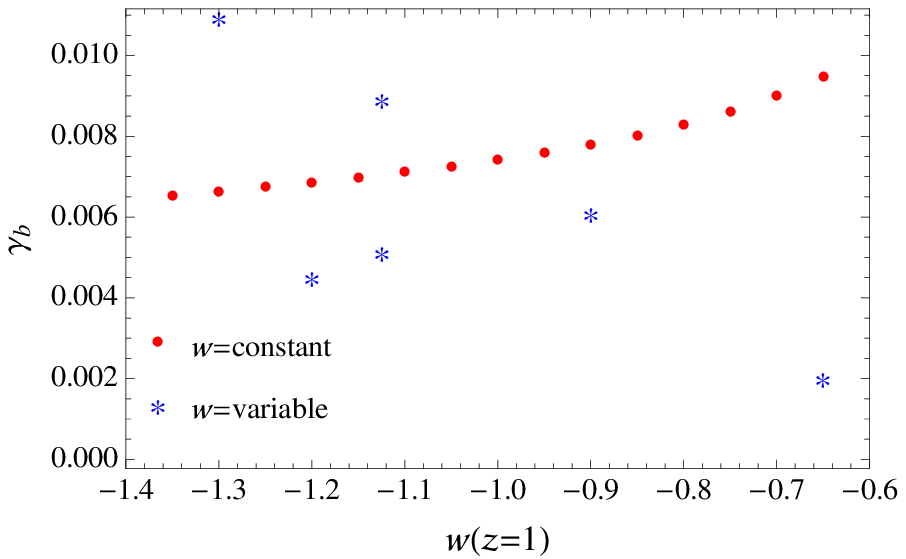}}\\ \hline
\end{tabular}
\caption{\label{f:gamma} 
We plot the best fit values for the growth index parameters as a function of $w$ evaluated at $z=1$. Because of this, it is possible for two models to fall at the on the same line vertically in our graphs, as some indeed do.  TOP:  We fit a constant $\gamma$ to our obtained logarithmic growth rate $f$ via the usual ansatz $f(z)=\Omega_m(z)^\gamma$.  Interestingly, these best fit values as a function of the dark energy equation of state, $w$, follow a linear trend.  We plot the best fit trend as a function of $w$, for which we find $\gamma = 0.552 + 0.028(1+w(z=1))$.  BOTTOM LEFT: We plot the best fits for the parameter $\gamma_e$ from the parameterization for $\gamma$ given by Eq. (\ref{eq:gammaexp}).  BOTTOM RIGHT:  We plot the best fits for the parameter $\gamma_b$ from the parameterization for $\gamma$ given by Eq. (\ref{eq:gammaexp}).  Notice all of the values remain positive even when DE perturbations have been introduced. This is consistent with previous results in absence of DE perturbations that showed this parameter takes positive values for dark energy models but can be negative for modified gravity models \cite{PolarskiDEP,WuDEP, DossettDEP}.} 
\end{figure}

Our results for these fits can be found in Fig. \ref{f:gamma}.  Notice that, for a constant gamma, the best fit values follow a mostly linear trend with respect to the dark energy equation of state, $w$, evaluated at a redshift, $z=1$.  We fit this trend and found the following relation
\be
\gamma = 0.552 + 0.028(1+w(z=1)).
\label{eq:gammaexp}
\ee
This is in agreement with the relation found by \cite{Linder2005DEP}.  In that work it was assumed that dark energy did not have perturbations and $w$ only affected the growth via its contribution to the Hubble expansion terms in the growth equations.  Here we have allowed dark energy perturbations and, for the scales at which we have evaluated $f$, their effect on the growth index is minimal.  Thus as a first main finding, the growth index, $\gamma$, remains a valid way to distinguish between different models of gravity even when dark energy is allowed to have density and velocity perturbations.  None of the values obtained for $\gamma$ are near those seen for modified gravity models such as the Dvali-Gabadadze-Porrati (DGP) model \cite{DGPDEP} or the $f(R)$ models  (\textit{e.g.} $\gamma_{DGP} = 0.6875$ \cite{LC07DEP} and $\gamma_{f(R)} = 0.42$ \cite{GMPDEP,MotohashiDEP}).

Another thing worth noticing is the consistent sign of the best fit values for the parameter $\gamma_b$.  This is consistent with the conclusions made in \cite{PolarskiDEP,WuDEP,DossettDEP}, where it was noted that the sign of slope parameter of the growth index (in this case $\gamma_b$) could be used to discriminate between different models of gravity. For models where GR is the underlying gravity theory \cite{DossettDEP} found a positive $\gamma_b$. We continue to see this trend.

\subsubsection{Impact on the MG parameter $Q$ \label{sec:POmgq}}

\begin{figure}[t!]
\centering
\begin{tabular}{|c|c|}
\hline 
{\includegraphics[width=3.45in,height=1.92in,angle=0]{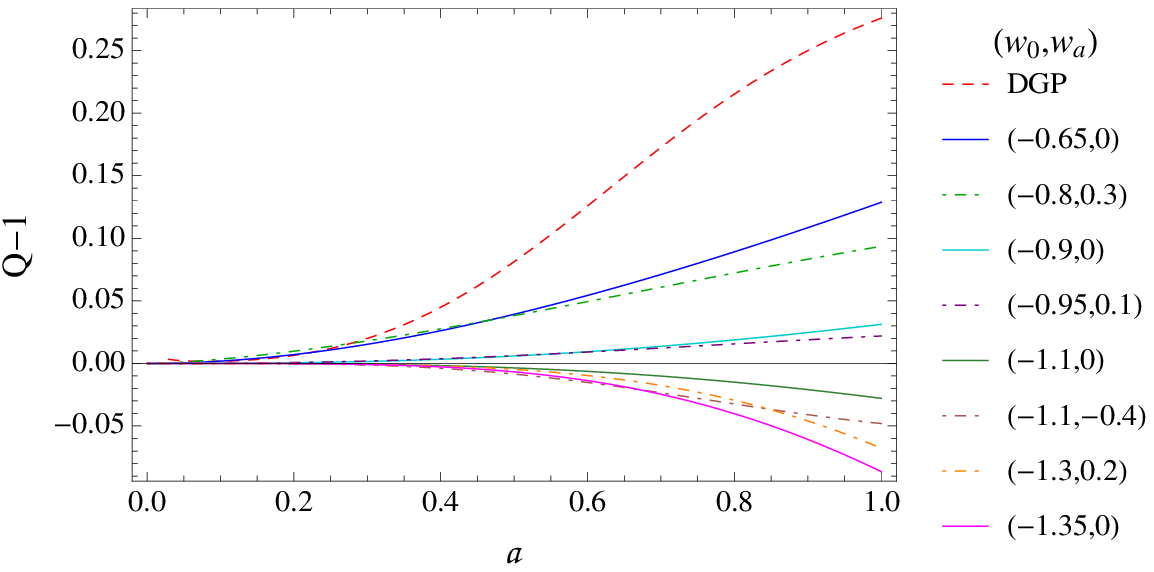}} &
{\includegraphics[width=3.45in,height=1.92in,angle=0]{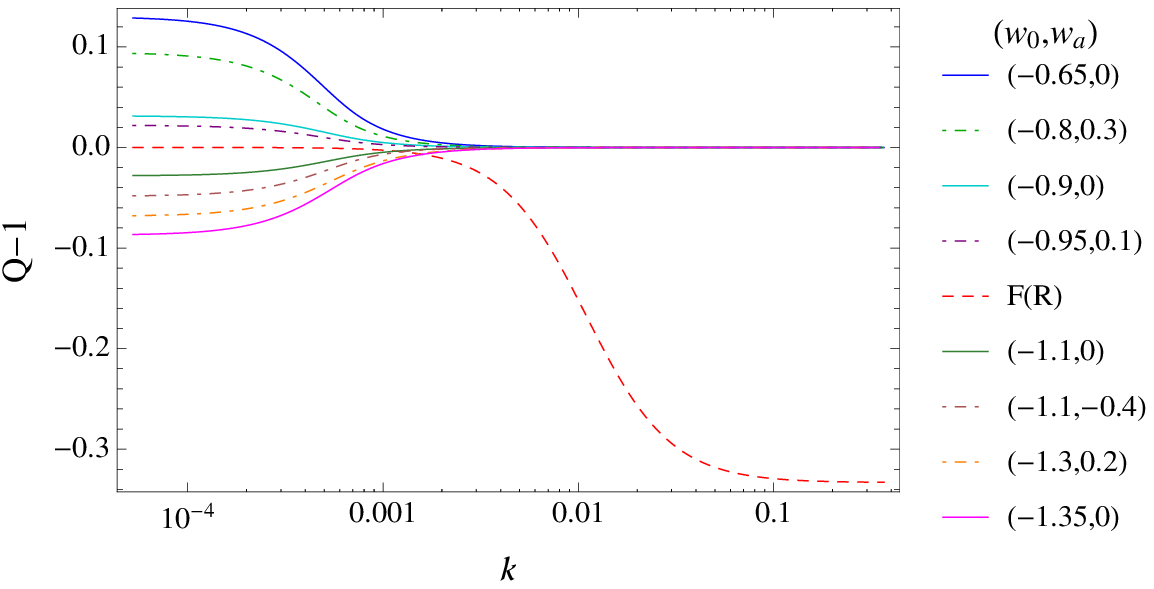}}\\ \hline
\end{tabular}
\caption{\label{f:QDE} 
We plot $Q-1$ as evaluated from Eq. (\ref{eq:Qeval}) for various dark energy models where we allow for dark energy perturbations.  The legend lists the various models used and is ordered according to their values at large scales today.  LEFT: Here we plot $Q-1$ as a function of scale factor $a$ for the scale corresponding to 40 times that of the horizon as explained in subsection (IV.A.2).  We also include a plot of $Q-1$ for a DGP model with $\Omega_m = 0.251$ and an expansion history identical to that of $\Lambda$CDM, as given by Eq. (\ref{eq:QDGP}).  This model deviates much more significantly than do any of the dark energy models with perturbations.  RIGHT: We plot $Q-1$ as a function of wave-number $k$ today.  Also included is a plot of $Q-1$ for a $f(R)$ model as described by Eq (\ref{eq:QFR}). Not only is the deviation that manifests for $f(R)$ more significant than the deviations for dark energy models with perturbations, but $f(R)$ also shows deviations for a different range of $k$ values. } 

\end{figure}

In order to explore how dark energy perturbations affect $Q$ we allow our modified version of \texttt{CAMB} to evolve dark energy perturbations for a range of values for the dark energy equation of state, $w$.  At each time step we then output $k$, $a$, and $Q$, where $Q$ is evaluated at each time step using Eq. (\ref{eq:Qeval}).  In Fig. \ref{f:QDE} we plot $Q-1$ as a function of $k$ today and as a function of $a$ for at 40 times the horizon scale.  We chose to plot the values of $Q-1$ as a funciton of $a$ at this scale because we found that larger scale modes did not significantly contribute (above a percent level) to the amplitude of even the lowest multipoles of the CMB power spectra.  In this way we would never be able detect variations in the MG parameters at scales larger than this.  

For comparison, in the plot of $Q-1$ as a function of $k$ we also plot the $Q-1$ for a $f(R)$ model using the of the parameterization of \cite{GMSM09}, which is an improved version of what was introduced by \cite{BZ08DEP}. In this parameterization, $Q$ is written as
\be
Q_{{f(R)}} = \frac{1}{1-1.4\times 10^{-8}\abs{\lambda_1}^2a^3}\frac{1+\frac{2}{3}\lambda_1^2k^2a^{4}}{1+\lambda_1^2k^2a^{4}},
\label{eq:QFR}
\ee
where $\lambda_1$ is just the Compton wavelength today. We can write $\lambda_1^2 = B_0c^2/(2H_0^2)$ and use $B_0$ to quantify the value of $\lambda_1$ in units of the Hubble radius. Thus the effect of $f(R)$ models on the growth is described by only 1 parameter, $B_0$. Here we plot an $f(R)$ model with $B_0 = 10^{-3}$ which is two orders of magnitude smaller than the upper limits placed on this parameter by \cite{GMSM09}.

When plotting $Q-1$ is a function of $a$, we include a plot of $Q-1$ for a DGP model with an expansion history matching that of $\Lambda$CDM with $\Omega_m = 0.251$.  $Q$ for a DGP model is given by \cite{Linder11DEP}
\be
Q_{DGP} = \frac{4+2\Omega_m(a)^2}{3+3\Omega_m(a)^2}.
\label{eq:QDGP}
\ee

Figure \ref{f:QDE} shows that for low $k$ values, $Q$ can indeed deviate from the GR value of 1 if dark energy is allowed to have perturbations.  These deviations are, however, nowhere close to the magnitude of deviations that appear for the $f(R)$ model shown -- even for the most extreme values of $w$.  On top of that, the deviations from the $f(R)$ model appear for a different range of $k$. The $f(R)$ model shows deviations for $k>0.003$ while the dark energy models show deviations for $k<0.003$.  The plot for $Q-1$ as a function of $a$ shows deviations are also not of the magnitude of those exhibited by the given DGP model where $Q$ is given by Eq. (\ref{eq:QDGP}).  Plugging in $\Omega_m = 0.251$ to this formula gives a value of $Q_{DGP}-1 = 0.294$. Again, this is well outside of the values shown in the plots -- even for the most extreme values of the dark energy equation of state. 

As discussed in Section \ref{sec:MGderiv} the MG parameter $R$ will not be greatly affected by the presence of dark energy perturbations that do not include anisotropic stress perturbations.  We did check to verify this and found that $R$ deviates from one by maximum of only $R-1\sim 3\times10^{-5}$.  In the next section, where we discuss dark energy models which have anisotropic stress perturbations, deviations of $R$ from unity can be much more pronounced.  

As an aside, it is worth mentioning that the modification to Einstein gravity proposed in \cite{CuscutonDEP} produces models with $R=1$ while allowing for $Q$ to be different from $1$. For a given potential for such models, one will need to derive the predicted parameter $Q$ and compare it to observations. The comparative constraints on the \textit{effective} equation of state of such models and the $Q$ parameter should provide a direct way to test such models. For example, the quadratic potential discussed there \cite{CuscutonDEP} will have an effective equation of state equivalent to that of a $\Lambda$CDM while showing a $Q$ different from $1$. Since the $\Lambda$CDM model has $Q=1$ this will be enough to distinguish it from the quadratic model. Second, their exponential potential model (see \cite{CuscutonDEP}) has an \textit{effective} equation of state of a DGP model while a value of $1$ for the $R$-parameter. This is distinct from the DGP model case where $R$ is different from $1$. So again, the use the effective equation of state in conjunction with the MG parameters $Q$ and $R$ is a means to test this particular modification to Einstein gravity.

\subsection{Effects of dark energy models with anisotropic stress perturbations\label{sec:pertpi}}
We will now shift our attention to models of dark energy that also include anisotropic stress perturbations in addition to density and velocity perturbations.  We will use three different models for dark energy anisotropic stress, labeling them Model I, II, and III.  

Our first model of dark energy models with anisotropic stress is the one discussed in section \ref{sec:DEpi}.  For the models we study we set $c_{\rm vis}^2 = 0.1$.  This value of $c_{\rm vis}^2$ was chosen because it produces anisotropic stress perturbations that have a realistic (not oversized) magnitude compared to the  mass averaged overdensity, $\bar{\Delta} = \sum_i \rho_i \Delta_i/\sum_i \rho_i$.  

\begin{figure}[t!]
\centering
\begin{tabular}{|c|c|}
\hline 
{\includegraphics[width=3.45in,height=1.92in,angle=0]{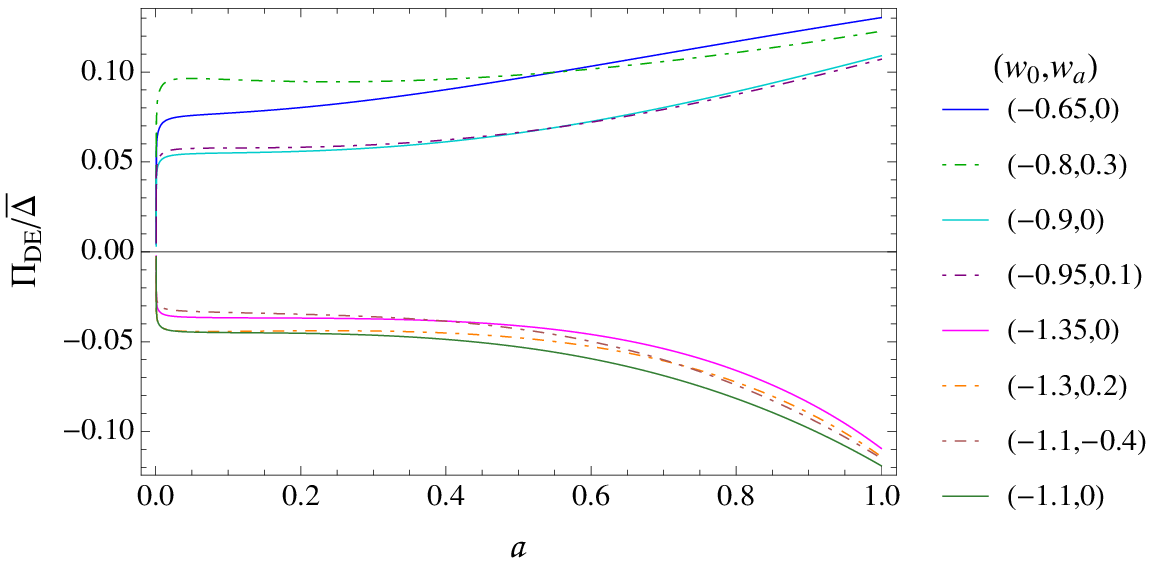}} &
{\includegraphics[width=3.45in,height=1.92in,angle=0]{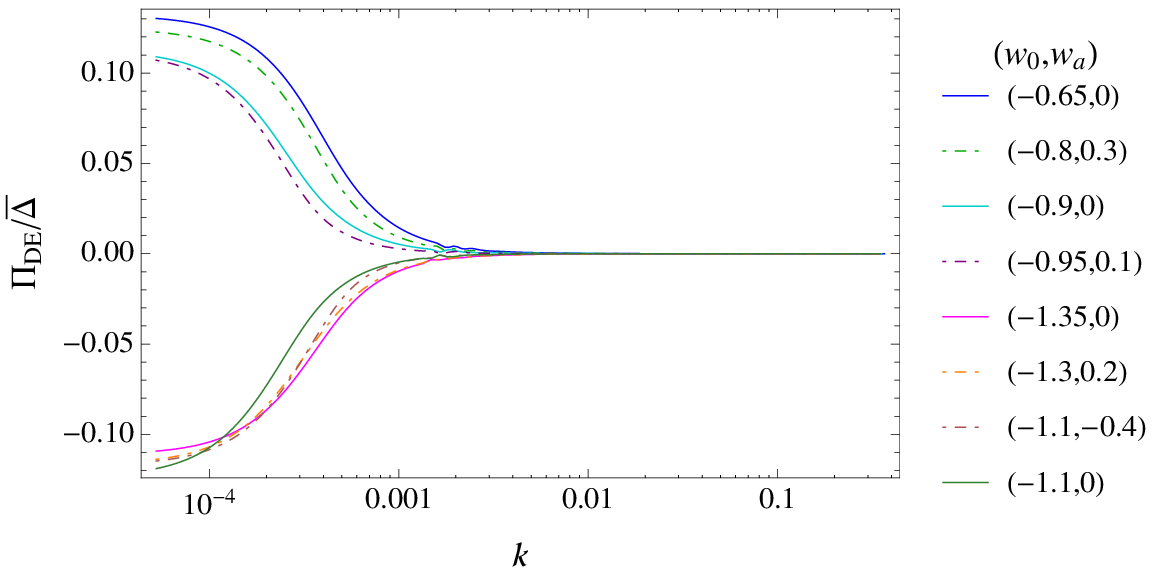}}\\ \hline
\end{tabular}
\caption{\label{f:DEpi} 
We plot $\Pi_{_{DE}}/\bar{\Delta}$ for Model I for various dark energy models.  For each plot, the legend lists the various models used and is ordered according to their values at large scales today. LEFT: We plot $\Pi_{_{DE}}/\bar{\Delta}$ as a function of scale factor $a$ on large scales. RIGHT: We plot $\Pi_{_{DE}}/\bar{\Delta}$ as a function of wave-number $k$ today.} 
\end{figure}

In Fig. \ref{f:DEpi} we plot $\Pi_{_{DE}}/\bar{\Delta}$ for this model of anisotropic stress perturbations for various dark energy models.  We include plots of $\Pi_{_{DE}}/\bar{\Delta}$ versus wave-number, $k$, today, as well as scale factor, $a$. The plots as a function of $a$ are taken a scale corresponding to 40 times the horizon scale as explained in in subsection (IV.A.2).

In this and other figures for $\Pi_{_{DE}}$, we compare $\Pi_{_{DE}}$ to the mass averaged overdensity, $\bar{\Delta}$, because the anisotropic stress perturbation and the overdensity are the two perturbation quantities that contribute to the metric potentials.  The mass averaged overdensity is used because it gives provides a general scale with which to compare other perturbation quantities.

\begin{figure}[t!]
\centering
\begin{tabular}{|c|c|}
\hline 
{\includegraphics[width=3.45in,height=1.92in,angle=0]{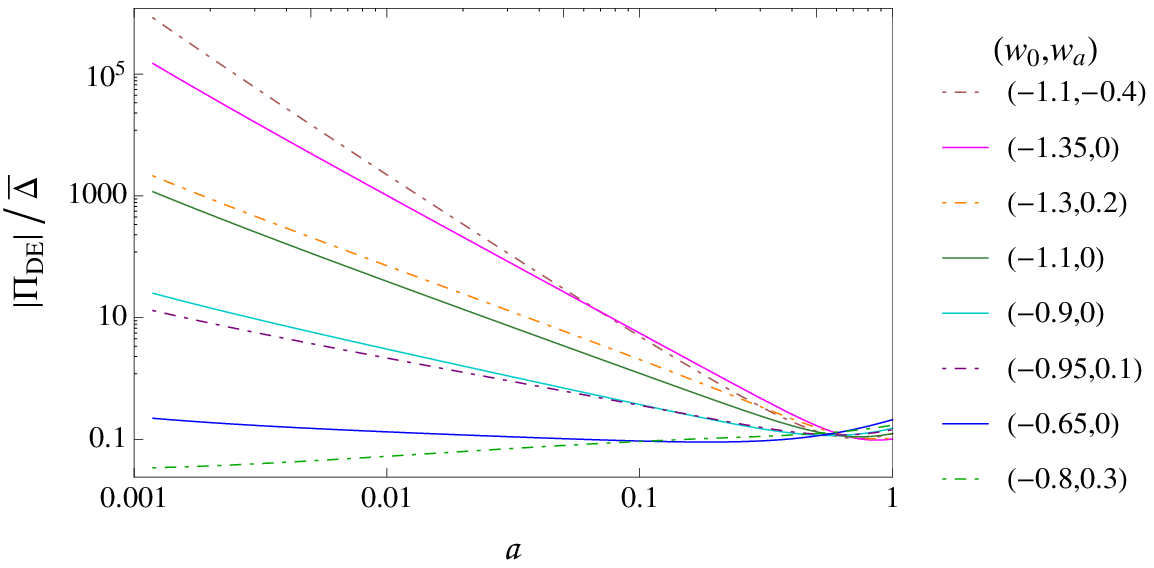}} &
{\includegraphics[width=3.45in,height=1.92in,angle=0]{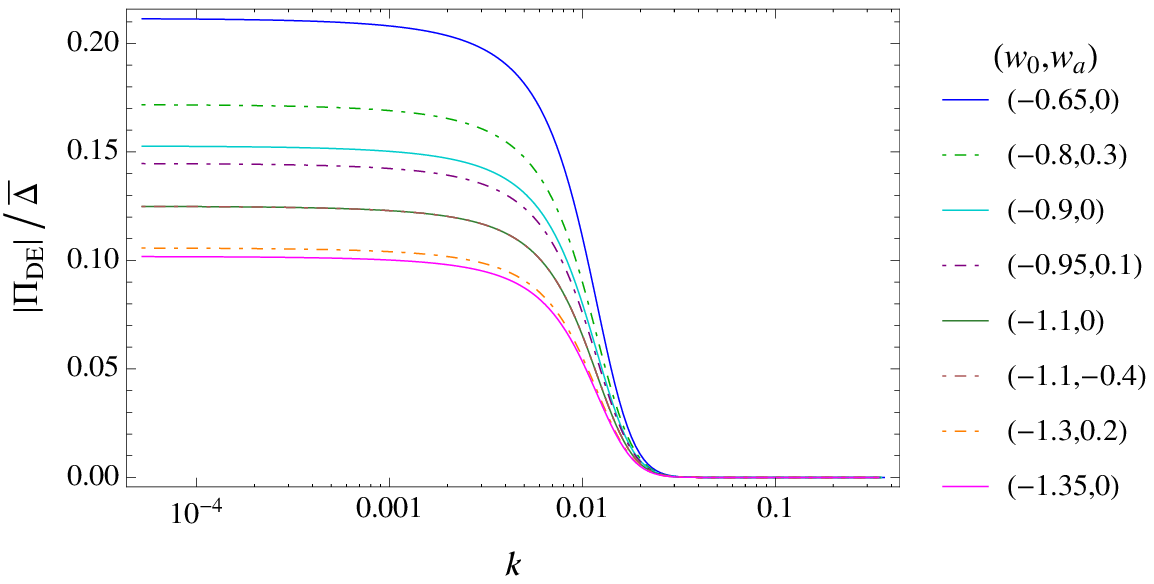}}\\ \hline
\end{tabular}
\caption{\label{f:Rpi} 
We plot $\abs{\Pi_{_{DE}}}/\bar{\Delta}$ for Models II and III as evaluated from Eq. (\ref{eq:DEpi}) for various dark energy models.  For each plot, the legend lists the various models used and is ordered according to their values at large scales today.  For Model II, we find $\Pi_{_{DE}}/\bar{\Delta}$ to be positive, while for Model III it is negative. LEFT: We plot $\abs{\Pi_{_{DE}}}/\bar{\Delta}$ as a function of scale factor $a$ on large scales. RIGHT: We plot $\abs{\Pi_{_{DE}}}/\bar{\Delta}$ as a function of wave-number $k$ today.} 

\end{figure}

In addition to considering anisotropic stress perturbations of Model I, we also look at dark energy models where the evolution and magnitude of $\Pi_{_{DE}}$ is such that it causes the MG parameter $R$ to behaves as:
\be
R(k,a) = \frac{1}{2}\Big[R_0-1+(1-R_0)\tanh 150(k-0.01)\Big]a^{1.8}+1.
\label{eq:MGR}
\ee
This form for $R$ was chosen by assuming that the time evolution of the parameter $R$ will be similar to that of the MG parameter $Q$ coming from the dark energy density perturbations in the dark energy models of the previous section.  As such, the time evolution of this model was chosen by fitting the evolution of $Q$ for the  $(w_0,w_a) = (0.65,0)$ dark energy model shown in Fig. \ref{f:QDE}.  We have chosen the wave-number $k$ where $R$ transitions to 1 to be $k = 0.01$, which is larger than what was found for $Q$ and is also larger than that which is observed for the dark energy model where $\Pi_{_{DE}}$ is evolved according to Eq. (\ref{eq:pidot}).  This allows us to consider more exotic dark energy anisotropic stress perturbations, thereby allowing us to make more concrete conclusions about the ability of dark energy perturbations to affect constraints on the MG parameters. 

From the behavior of $R$ that we have chosen, we can use Eq. (\ref{eq:Reval}) and solve for $\Pi_{_{DE}}$.   To avoid some numerical instabilities, we assume the evolution for $R$ described by Eq. (\ref{eq:MGR}) comes only from the contribution of $\Pi_{_{DE}}$ and, for this reason, ignore the anisotropic stress perturbations of other matter species when calculating $\Pi_{_{DE}}$.  Solving this equation gives
\be
\Pi_{_{DE}} = \frac{(R-1)}{2}\frac{\rho_{_{DE}}\Delta_{_{DE}}+\sum\limits_{i\ne DE}\rho_i\Delta_i}{\rho_{_{DE}}w_{_{DE}}}.
\label{eq:DEpi}
\ee 

Now that we are able to calculate $\Pi_{_{DE}}$ for a given $R$ we can sub this quantity into the dark energy perturbation evolution equations given by Eqs. (\ref{eq:DEdeltadot}) and (\ref{eq:DEthetadot}) so the dark energy perturbations are evolved consistent with the presence of such a $\Pi_{_{DE}}$. 

Here we will consider models where $R_0$ of $0.8$ and $1.2$ and refer to these models of $\Pi_{_{DE}}$ as Model II and Model III respectively.  In Fig. \ref{f:Rpi}, we plot behavior of $\Pi_{_{DE}}/\bar{\Delta}$ for these.  This figure includes plots of $\Pi_{_{DE}}/\bar{\Delta}$ versus wave-number, $k$, today, as well as scale factor, $a$. The plots as a function of $a$ are taken considering scales of 40 times the horizon scale as explained in subsection (IV.A.2).  

As expected from our chosen form for $R$, at smaller scales the anisotropic stress perturbation is negligible with respect to the mass averaged overdensity.  On larger scales, however, the $\Pi_{_{DE}}$ required to cause $R$ to deviate from one approaches the same order of magnitude as the $\bar{\Delta}$. From the time evolution plots we see that, at early times, a very large $\Pi_{_{DE}}$, compared to $\bar{\Delta}$, is required to cause $R$ to depart even slightly from unity.  This is due to the fact that, at early times, the dark energy density is very, very small and the dark energy anisotropic stress perturbation must make up for this in order to influence $R$ in any way.  This can be seen in Eq. (\ref{eq:Reval}).

Now that we have described the three models of dark energy anisotropic stress perturbations, we can discuss how dark energy models with anisotropic stress perturbations will affect the various parameters we are using to test gravity. 

\subsubsection{Impact on the growth index \label{sec:PPgi}}
We will again start by discussing the effect of the dark energy perturbations on the growth index $\gamma$, and parameters from its redshift dependent parameterization given by Eq. (\ref{eq:gammaexp}).  Interestingly, for Model I, we obtain results identical to those obtained for dark energy models without any anisotropic stress perturbations.  The main reason for this is that, as can be seen in Fig. \ref{f:DEpi}, the anisotropic stress perturbations are zero for the scale where we are evaluating $f$ ($k = 0.02$) and thus do not affect the growth of perturbations on those scales.  For this reason, the discussion in this subsection will mainly be focused on models II and III, which, as can be seen in Fig. \ref{f:Rpi}, do have anisotropic stress perturbations at the scales we consider when evaluating $f$.

\begin{figure}[t!]
\centering
\begin{tabular}{|c|c|}
\hline
{\includegraphics[width=3.45in,height=1.92in,angle=0]{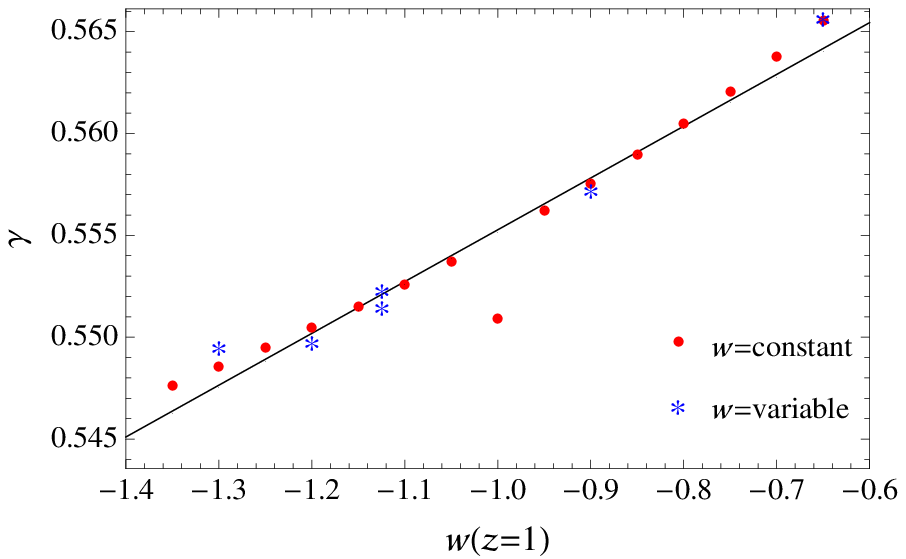}} &
{\includegraphics[width=3.45in,height=1.92in,angle=0]{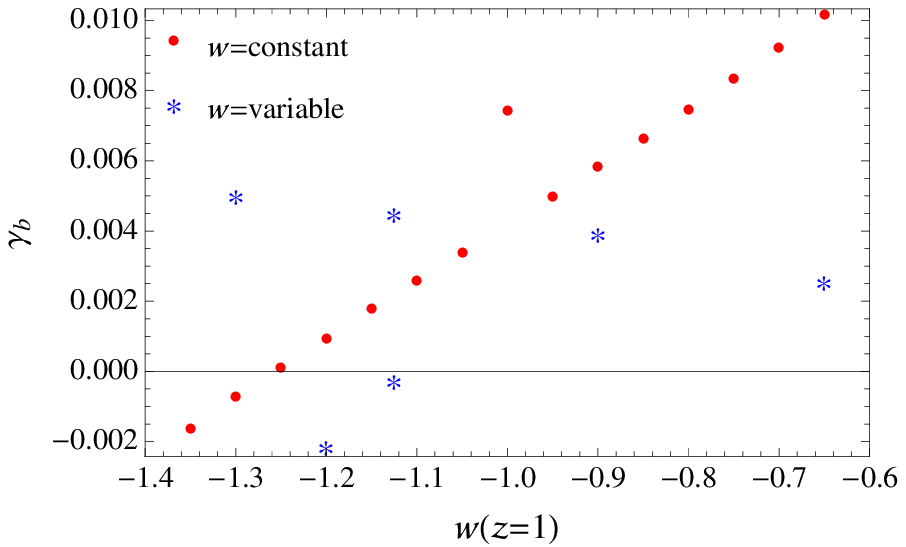}}\\ \hline
\end{tabular}
\caption{\label{f:Rp8gamma} 
We plot the best fit values for the growth index parameters as a function of $w$ evaluated at $z=1$ for anisotropic stress model II.  LEFT: We fit a constant $\gamma$ to our obtained logarithmic growth rate $f$ via the usual ansatz $f(z)=\Omega_m(z)^\gamma$.  Again, these best fit values as a function of the dark energy equation of state, $w$, follow a linear trend.  We plot the best fit trend as a function of $w$, for which we find $\gamma = 0.555 + 0.025(1+w(z=1))$.  This corresponds to a shift towards higher values of $\gamma$ compared to models without anisotropic stress. RIGHT:  We plot the best fits for the parameter $\gamma_b$ from the parameterization for $\gamma$ given by Eq. (\ref{eq:gammaexp}).  All of the values for the parameter are shifted towards lower values compared to the case where there was no anisotropic stress.  This even causes some of the values to go negative.  This unfortunately shows that the sign of this parameter is not a consistent way to distinguish between dark energy and modified gravity models.} 
\end{figure}

The effect of the dark energy anisotropic stress perturbations from model II on $\gamma$ and $\gamma_b$ are shown in Fig. \ref{f:Rp8gamma}.  We once again plot the best fit values obtained as described in section \ref{sec:POgi} vs. $w(z=1)$.  For brevity we do not plot $\gamma_e$ as the effect of $\Pi_{_{\rm DE}}$ is similar to what is seen for $\gamma$.  The first thing to notice is that all of the values for $\gamma$ are shifted towards higher values.  This is reflected in the fact that the best fit linear trend for $\gamma$ as a function of $w(z=1)$ is given by 
\be
\gamma = 0.555 + 0.025(1+w(z=1)).
\label{eq:Rp8gammaexp}
\ee
While the values of $\gamma$ are shifted up, the shift corresponds to a difference of only about $0.5\%$ of its value for dark energy models where $\Pi_{_{\rm DE}} = 0$.  These values are still well away from values obtained in the DGP or $f(R)$ modified gravity models as described previously in section \ref{sec:POgi}.   

A more significant result though is the fact that for some models the value of $\gamma_b$ goes negative.  This is significant because, as discussed previously, the sign of the slope parameter of the growth index has been thought to be a good way to discriminate between dark energy and modified gravity.  For this model of dark energy with anisotropic stress, the previous trend no longer holds and thus the sign of $\gamma_b$ or any of the parameters relating to the redshift slope of the growth index cannot be said to be robust to dark energy models with anisotropic stress perturbations.  

\begin{figure}[t!]
\centering
\begin{tabular}{|c|c|}
\hline
{\includegraphics[width=3.45in,height=1.92in,angle=0]{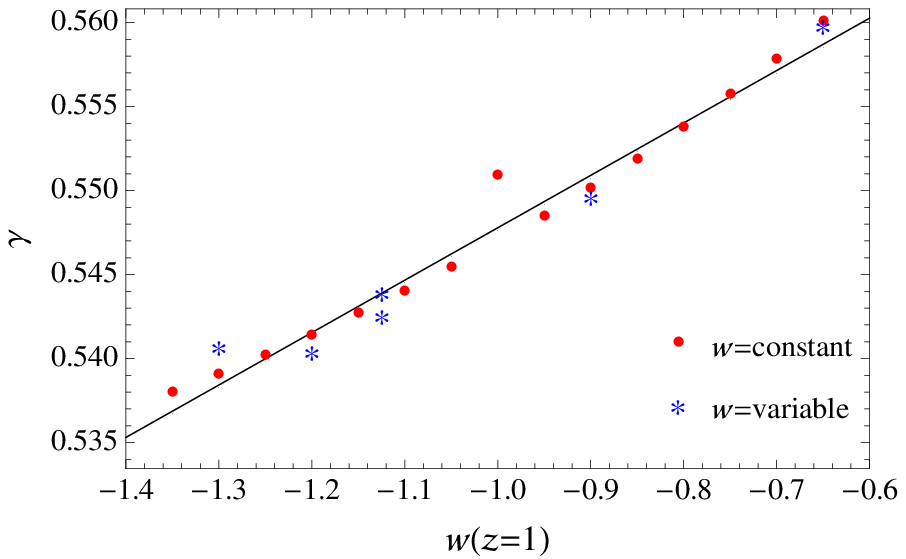}} &
{\includegraphics[width=3.45in,height=1.92in,angle=0]{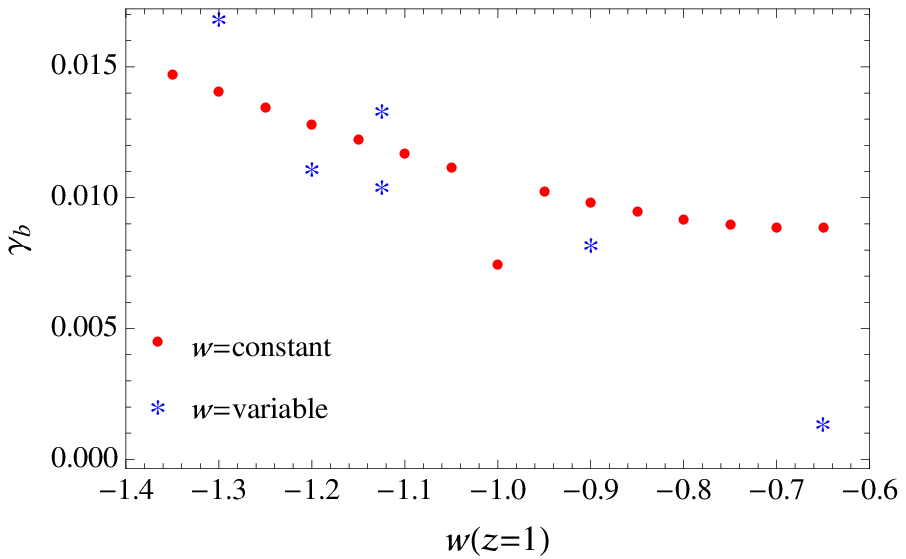}}\\ \hline
\end{tabular}
\caption{\label{f:R1p2gamma} 
We plot the best fit values for the growth index parameters as a function of $w$ evaluated at $z=1$ for anisotropic stress model III.  LEFT: We fit a constant $\gamma$ to our obtained logarithmic growth rate $f$ via the usual ansatz $f(z)=\Omega_m(z)^\gamma$.  Again, these best fit values as a function of the dark energy equation of state, $w$, follow a linear trend.  We plot the best fit trend as a function of $w$, for which we find $\gamma = 0.548 + 0.031(1+w(z=1))$.  This corresponds to a shift towards lower values of $\gamma$ compared to models without anisotropic stress. TOP RIGHT:  We plot the best fits for the parameter $\gamma_b$ from the parameterization for $\gamma$ given by Eq. (\ref{eq:gammaexp}).  All of the values for the parameter are shifted towards higher values compared to the case where there was no anisotropic stress.  Unlike Model II in this model the values of this parameter stay positive.} 
\end{figure}

For dark energy models with anisotropic stress perturbations described by model III, the effect on $\gamma$ is opposite of what it was for model II. This can be seen in Fig. \ref{f:R1p2gamma} were we plot the best fit $\gamma$ and $\gamma_b$ vs. $w(z=1)$ for various dark energy models.  In this case the values of $\gamma$ are shifted towards slightly lower values. For these models the best fit linear trend of $\gamma$ as a function of $w(z=1)$ is given by 
\be
\gamma = 0.548 + 0.031(1+w(z=1)).
\label{eq:R1p2gammaexp}
\ee
Once again this shift is relatively small, corresponding to a change of only about $0.7\%$ compared to the values of $\gamma$ obtained for dark energy models where $\Pi_{_{\rm DE}} = 0$.  This combined with the results obtained for models I and II, as well as the results obtained in section \ref{sec:POgi} show that the growth index, $\gamma$ is indeed quite robust to the presence of all types of dark energy perturbations as a way to distinguish between dark energy and modified.  

The values obtained for $\gamma_b$ for these models do not go negative as they did in model II. These results combined with those obtained for model II are interesting and show that the sign of the redshift slope parameters of the growth index is not robust to the presence of dark energy models with anisotropic stress perturbations.

\subsubsection{Impact on the MG parameter $Q$ \label{sec:PPmgq}}
Next we will discuss the effect of dark energy models which include anisotropic stress perturbations on the MG parameter $Q$.  All three models of dark energy anisotropic stress perturbations have an effect on this parameter.  

\begin{figure}[t!]
\centering
\begin{tabular}{|c|c|}
\hline 
{\includegraphics[width=3.45in,height=1.92in,angle=0]{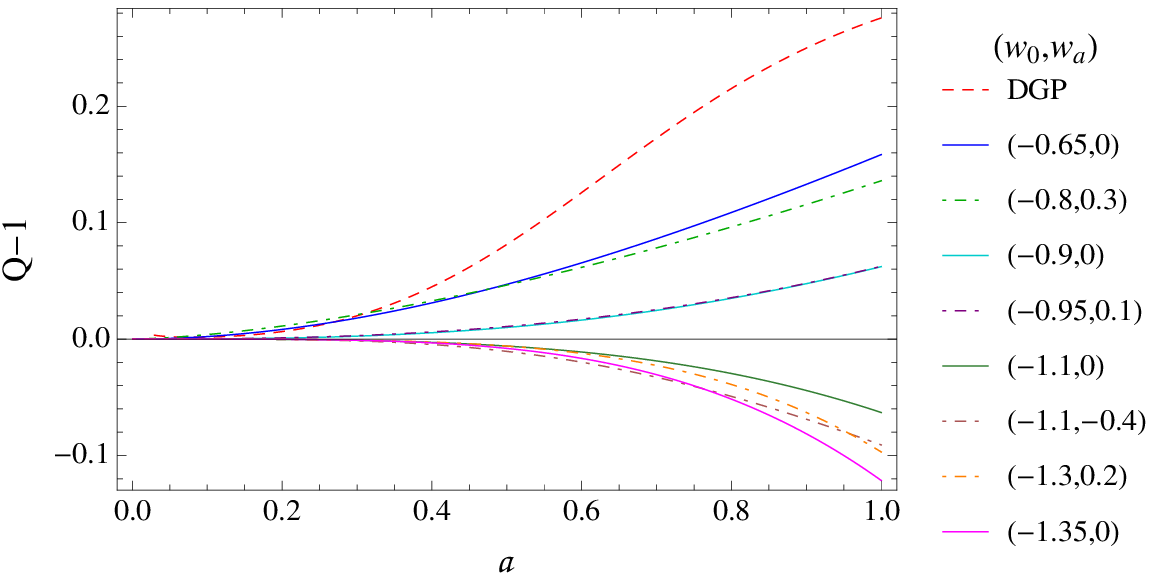}} &
{\includegraphics[width=3.45in,height=1.92in,angle=0]{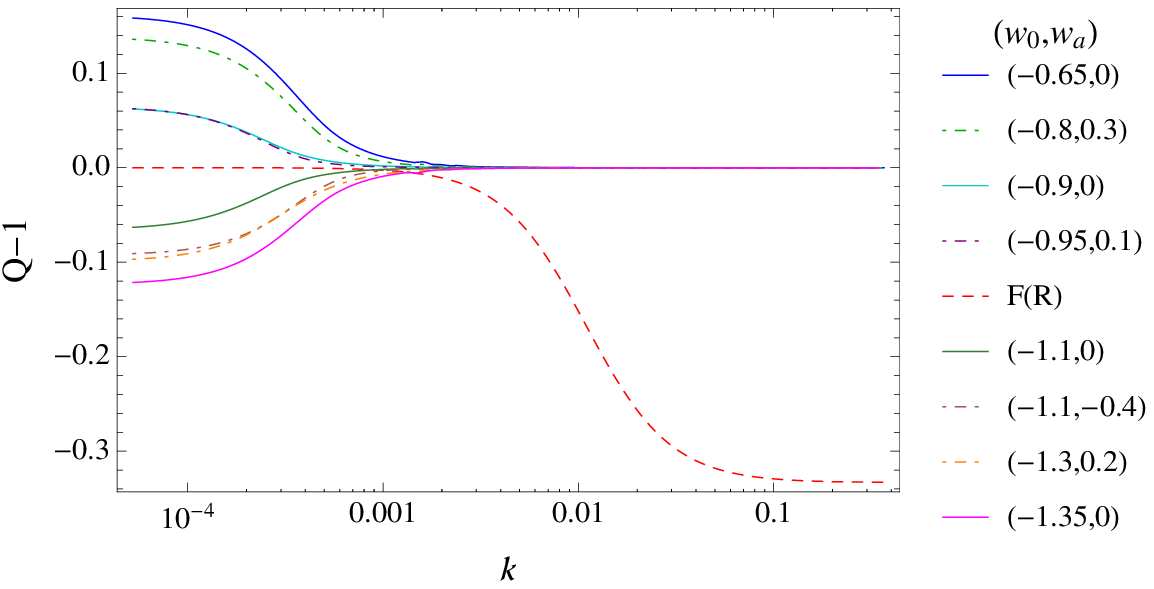}}\\ \hline
\end{tabular}
\caption{\label{f:DEpiQ} 
We plot $Q-1$ as evaluated from Eq. (\ref{eq:Qeval}) for various dark energy models where we allow for dark energy perturbations and model dark energy anisotropic stress as Model I.  The legend lists the various models used and is ordered according to their values on large scales today.  LEFT: Here we plot $Q-1$ as a function of scale factor $a$ for the scale corresponding to 40 times that of the horizon as explained in subsection (IV.A.2).  We also include a plot of $Q-1$ for a DGP model with $\Omega_m = 0.251$ and an expansion history identical to that of $\Lambda$CDM, as given by Eq. (\ref{eq:QDGP}).  This model deviates much more significantly than do any of the dark energy models with perturbations.  RIGHT: We plot $Q-1$ as a function of wave-number $k$ today.  Also included is a plot of $Q-1$ for a $f(R)$ model as described by Eq (\ref{eq:QFR}). Not only is the deviation that manifests for $f(R)$ more significant than the deviations for dark energy models with perturbations, but $f(R)$ also shows deviations for a different range of $k$ values. } 
\end{figure}

In Fig.\ref{f:DEpiQ} we plot the effect of dark energy models with anisotropic stress perturbations described by model I on $Q$.  Comparing these plots to those of Fig.\ref{f:QDE} we can see that the only difference is that the absolute value of the $Q$ for the various dark energy equations of state is larger at the largest scales considered.  The time and scale dependence of the MG parameter, though does not change much.  While the values of $Q$ exhibited by the dark energy models considered here is larger those without anisotropic stress perturbations, it still does not compare to the values predicted for various modified gravity models such as the DGP and $f(R)$ models.  Once again, for the $f(R)$ model in particular, the deviations exhibited by dark energy models and the modified gravity model occur in a completely different scale range.

\begin{figure}[t!]
\centering
\begin{tabular}{|c|c|}
\hline 
{\includegraphics[width=3.45in,height=1.92in,angle=0]{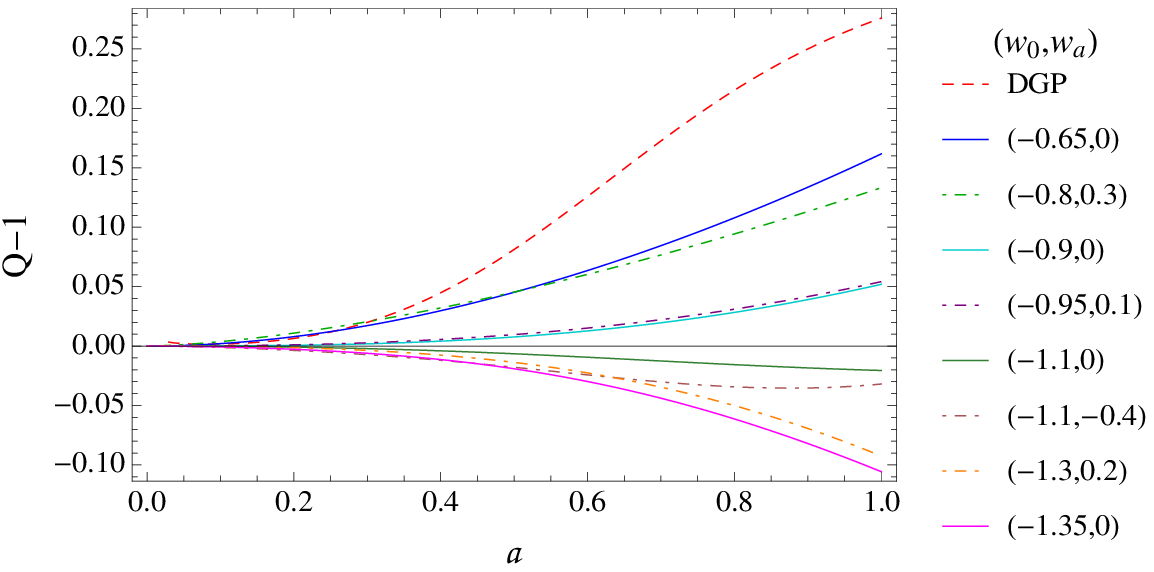}} &
{\includegraphics[width=3.45in,height=1.92in,angle=0]{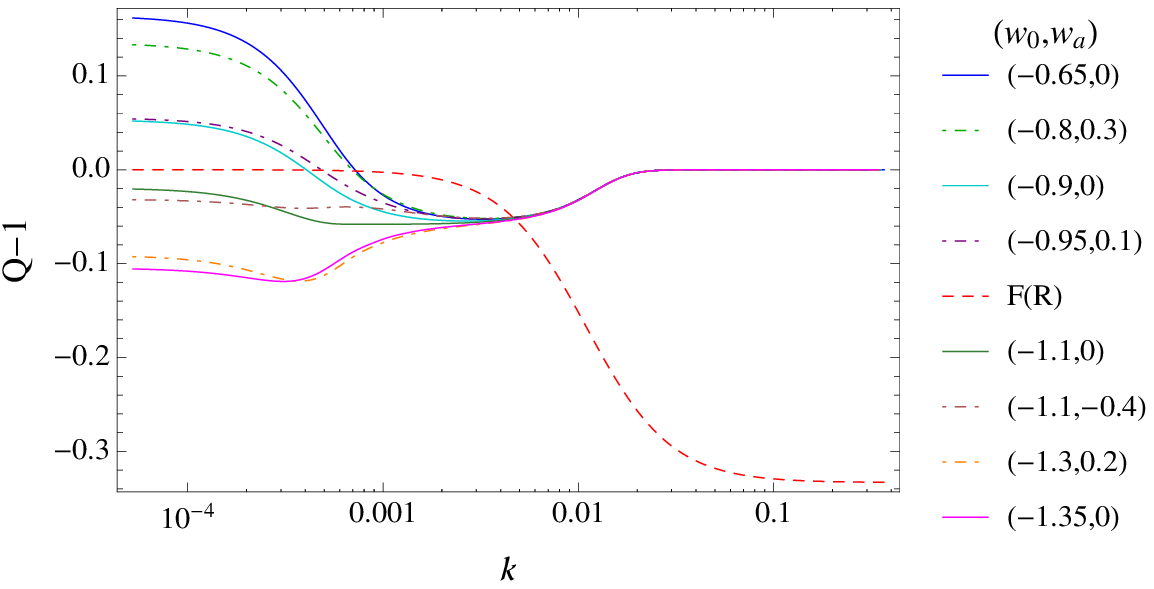}}\\ \hline
\end{tabular}
\caption{\label{f:Rp8Q} 
We plot $Q-1$ as evaluated from Eq. (\ref{eq:Qeval}) for various dark energy models where we allow for dark energy density and anisotropic stress perturbations as in Model II.  The legend lists the various models used and is ordered according to their values today.  LEFT: Here we plot $Q-1$ as a function of scale factor $a$ for the scale corresponding to 40 times that of the horizon as explained in subsection (IV.A.2).  We also include a plot of $Q-1$ for a DGP model with $\Omega_m = 0.251$ and an expansion history identical to that of $\Lambda$CDM, as given by Eq. (\ref{eq:QDGP}).  This model deviates much more significantly than do any of the dark energy models with perturbations.  RIGHT: We plot $Q-1$ as a function of wave-number $k$ today.  Also included is a plot of $Q-1$ for a $f(R)$ model as described by Eq (\ref{eq:QFR}). Not only is the deviation that manifests for $f(R)$ more significant than the deviations for dark energy models with perturbations, but $f(R)$ also shows deviations for a different range of $k$ values. } 

\end{figure}

\begin{figure}[t!]
\centering
\begin{tabular}{|c|c|}
\hline 
{\includegraphics[width=3.45in,height=1.92in,angle=0]{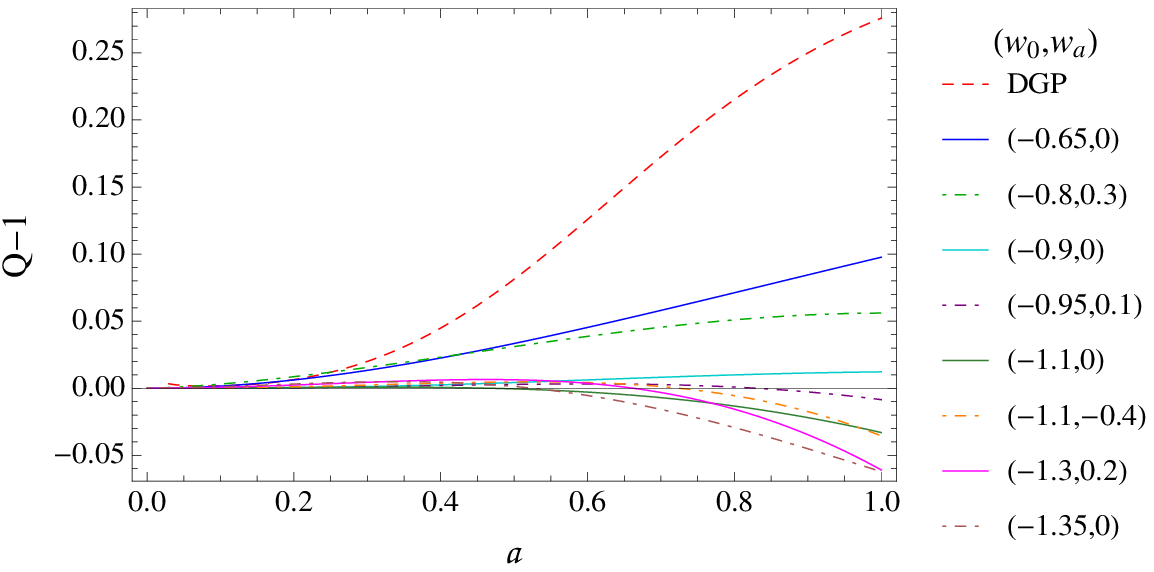}} &
{\includegraphics[width=3.45in,height=1.92in,angle=0]{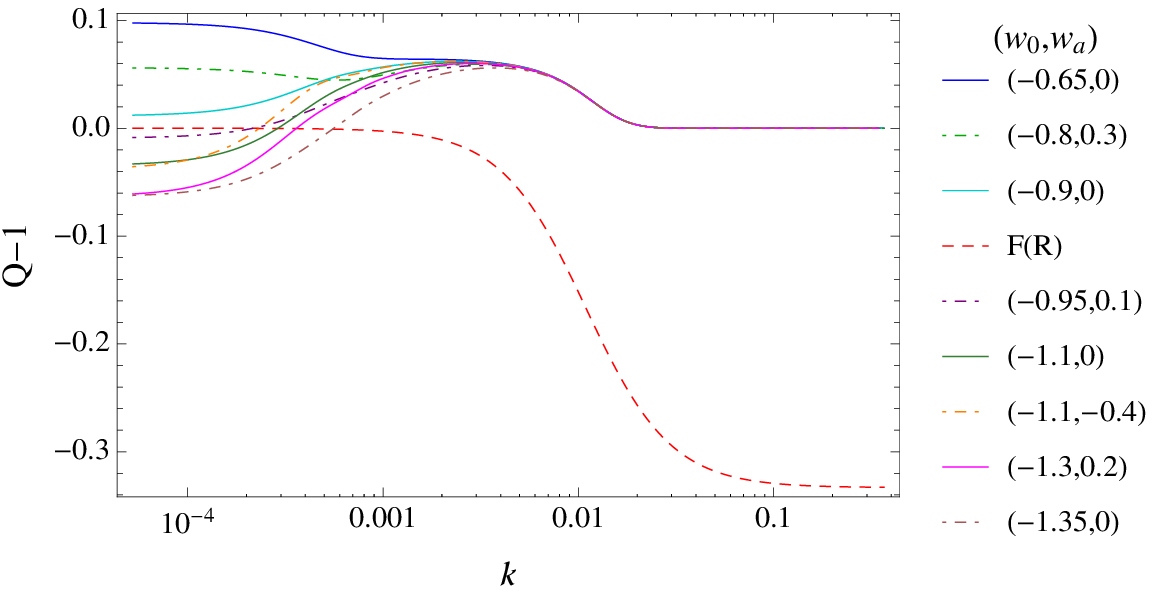}}\\ \hline
\end{tabular}
\caption{\label{f:R1p2Q} 
We plot $Q-1$ as evaluated from Eq. (\ref{eq:Qeval}) for various dark energy models where we allow for dark energy perturbations and model dark energy anisotropic stress as Model III.  The legend lists the various models used and is ordered according to their values today.  LEFT: Here we plot $Q-1$ as a function of scale factor $a$ for the scale corresponding to 40 times that of the horizon as explained in subsection (IV.A.2).  We also include a plot of $Q-1$ for a DGP model with $\Omega_m = 0.251$ and an expansion history identical to that of $\Lambda$CDM, as given by Eq. (\ref{eq:QDGP}).  This model deviates much more significantly than do any of the dark energy models with perturbations.  RIGHT: We plot $Q-1$ as a function of wave-number $k$ today.  Also included is a plot of $Q-1$ for a $f(R)$ model as described by Eq (\ref{eq:QFR}). Not only is the deviation that manifests for $f(R)$ more significant than the deviations for dark energy models with perturbations, but $f(R)$ also shows deviations for a different range of $k$ values. } 
\end{figure}

The effect of the models II and III on $Q$ is a little more subtle.  We plot this in figures \ref{f:Rp8Q} and \ref{f:R1p2Q} respectively.  In contrast to the smooth behavior of $Q$ observed in model I, these models cause $Q$ to vary a bit more widely.  For example, for these anisotropic stress models, at $k\sim 10^{-2}$ all of the dark energy models, regardless of their equation of state, have identical non-zero values for $Q-1$, and for half the models these values for $Q-1$ are of opposite sign compared to their values at larger scales.  Explaining this behavior can be done by looking at the comoving overdensity $\Delta_{\rm DE}$ because, as shown in Eq. (\ref{eq:Qeval}), $\Delta_{\rm DE}$ directly contributes to $Q-1$. Recall that there are are two components of $\Delta_{\rm DE}$, the density perturbation, $\delta_{\rm DE}$, and the quantity $3\mathcal{H}(1+w)\theta_{\rm DE}/k^2$ which of course is related to the velocity perturbation.  Looking at the behavior of these two components individually does indeed explain the behavior of $Q$ exhibited the dark energy models with anisotropic stress perturbations described by models II and III.  As discussed in \cite{KM05DEP}, a positive $\Pi_{_{\rm DE}}$ (model II) will act to enhance  the velocity perturbation and suppress the density perturbation.  The converse is true for models with a negative $\Pi_{_{\rm DE}}$ (model III). 

As it turns out the $\theta_{\rm DE}$ contribution to $\Delta_{\rm DE}$ is dominant at larger scales.  So as seen in Figs. \ref{f:Rp8Q} and \ref{f:R1p2Q} the large scale values of $Q-1$ are shifted up for model II (due to the enhancement of $\theta_{\rm DE}$) and, conversely, down for model III. However, at intermediate scales were all the dark energy models are seen to take on identical, non-zero values of $Q-1$, the dominant term in $\Delta_{\rm DE}$ is the density perturbation, $\delta_{\rm DE}$ and thus the value of $Q-1$ at $k \sim 10^{-2}$ is shifted, as seen, downwards for model II, and upwards for model III. 

In spite of the more complex behavior of $Q$ exhibited by dark energy models with anisotropic stress perturbations described by model II and III, we again see that the value of $Q$ for these models is not as large as the value for $Q$ the previously discussed modified gravity.  Once again, thought there is some overlap, the more significant deviations from unity in $Q$ for these dark energy models occur at a distinct scale range compared to those for the $f(R)$ modified gravity model.

\subsubsection{Impact on the MG parameter $R$ \label{sec:PPmgr}}
Now we can discuss the effect of the dark energy models with anisotropic stress perturbations on the MG parameter $R$. We first explore the effect of the model I $R$.  This is done by using Eq. \ref{eq:Reval}.  

\begin{figure}[t!]
\centering
\begin{tabular}{|c|c|}
\hline 
{\includegraphics[width=3.45in,height=1.92in,angle=0]{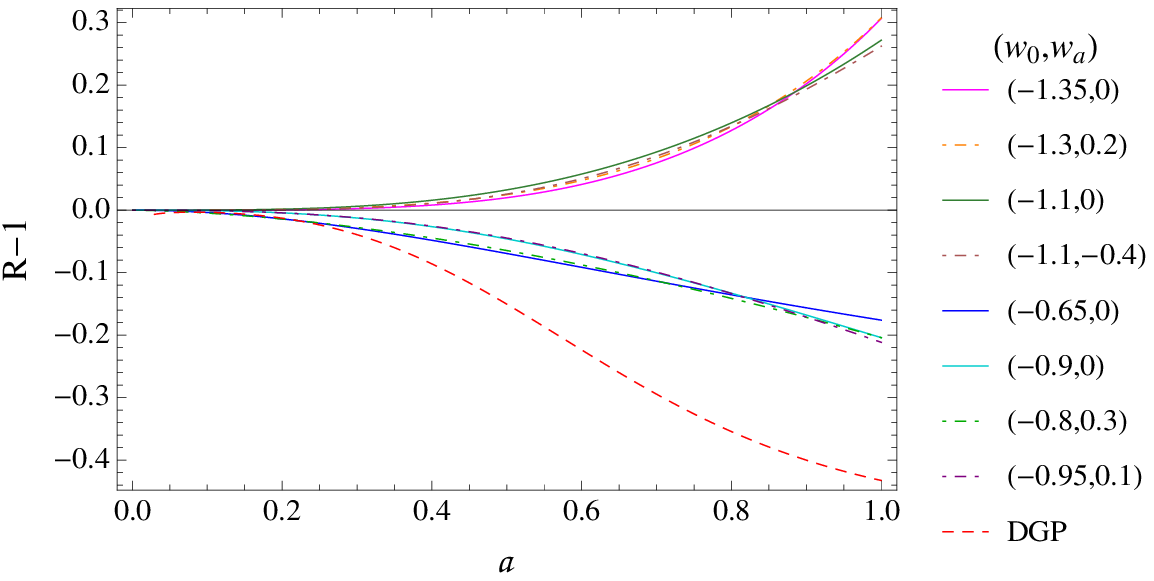}} &
{\includegraphics[width=3.45in,height=1.92in,angle=0]{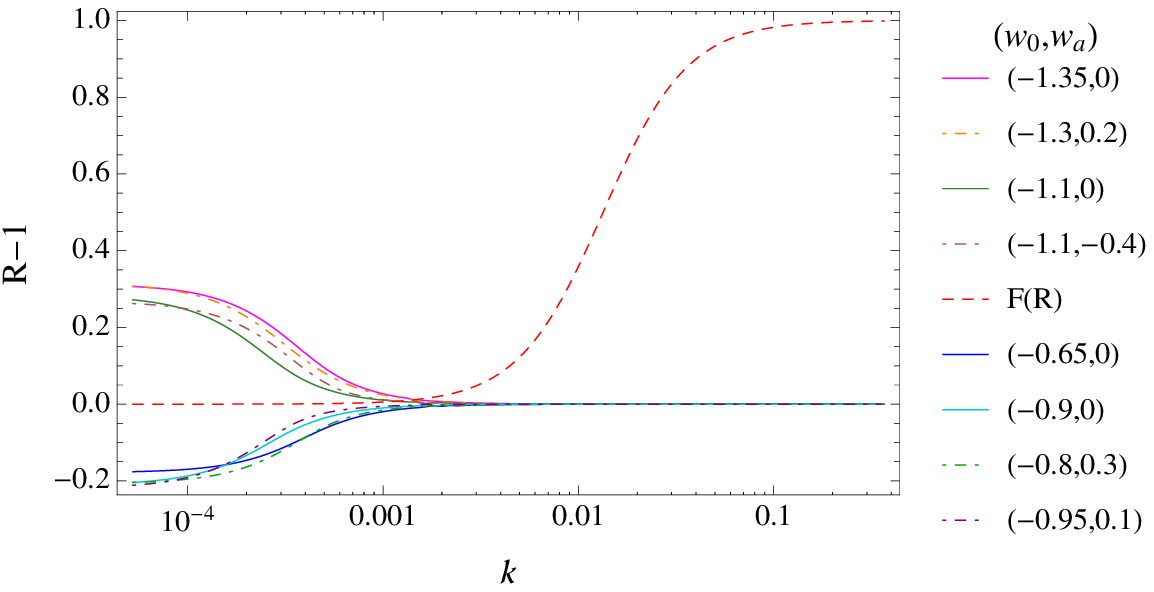}}\\ \hline
\end{tabular}
\caption{\label{f:DEpiR} 
We plot $R-1$ as evaluated from Eq. (\ref{eq:Reval}) for various dark energy models where we allow for dark energy perturbations and model dark energy anisotropic stress as Model I.  The legend lists the various models used and is ordered according to their values today.  LEFT: Here we plot $R-1$ as a function of scale factor $a$ for the scale corresponding to 40 times that of the horizon as explained in subsection (IV.A.2).  We also include a plot of $R-1$ for a DGP model with $\Omega_m = 0.251$ and an expansion history identical to that of $\Lambda$CDM, as given by Eq. (\ref{eq:RDGP}).  This model deviates much more significantly than do any of the dark energy models with perturbations.  RIGHT: We plot $R-1$ as a function of wave-number $k$ today.  Also included is a plot of $R-1$ for a $f(R)$ model as described by Eq (\ref{eq:RFR}). Not only is the deviation that manifests for $f(R)$ more significant than the deviations for dark energy models with perturbations, but $f(R)$ also shows deviations for a different range of $k$ values. } 
\end{figure}

Our results for model I are shown in Fig.\ref{f:DEpiR}.  For comparison, in the plot of $R-1$ as a function of $k$ we also plot the $R-1$ for a $f(R)$ model using the of the parameterization of \cite{GMSM09}, which is an improved version of what was introduced by \cite{BZ08DEP}. In this parameterization, $R$ is written as
\be
R_{{f(R)}} = \frac{1+\frac{4}{3}\lambda_1^2k^2a^{4}}{1+\frac{2}{3}\lambda_1^2k^2a^{4}},
\label{eq:RFR}
\ee
where as described previously, $\lambda_1$ is just the Compton wavelength today and can be written as $\lambda_1^2 = B_0c^2/(2H_0^2)$. Again, we plot an $f(R)$ model with $B_0 = 10^{-3}$ which is two orders of magnitude smaller than the upper limits placed on this parameter by \cite{GMSM09}.

When plotting $R-1$ is a function of $a$, we include a plot of $R-1$ for a DGP model with an expansion history matching that of $\Lambda$CDM with $\Omega_m = 0.251$.  For a DGP model, $R_{DGP}$, can be written as,\cite{Linder11DEP}
\be
R_{DGP}=\frac{1+2\Omega_m(a)^2}{2+\Omega_m(a)^2}.
\label{eq:RDGP}
\ee

As one can see the deviations in the value of $R$ from unity for these dark energy models is not as significant as the DGP model or the $f(R)$ model.  Similar to what was seen for the MG parameter $Q$, the $f(R)$ models exhibit deviations of $R$ from one that occur at completely different scales compared to the dark energy models. We can therefore conclude that dark energy models with anisotropic stress perturbations described by model I would be distinguishable from such modified gravity models.

\begin{figure}[t!]
\centering
\begin{tabular}{|c|c|}
\hline 
{\includegraphics[width=3.45in,height=1.92in,angle=0]{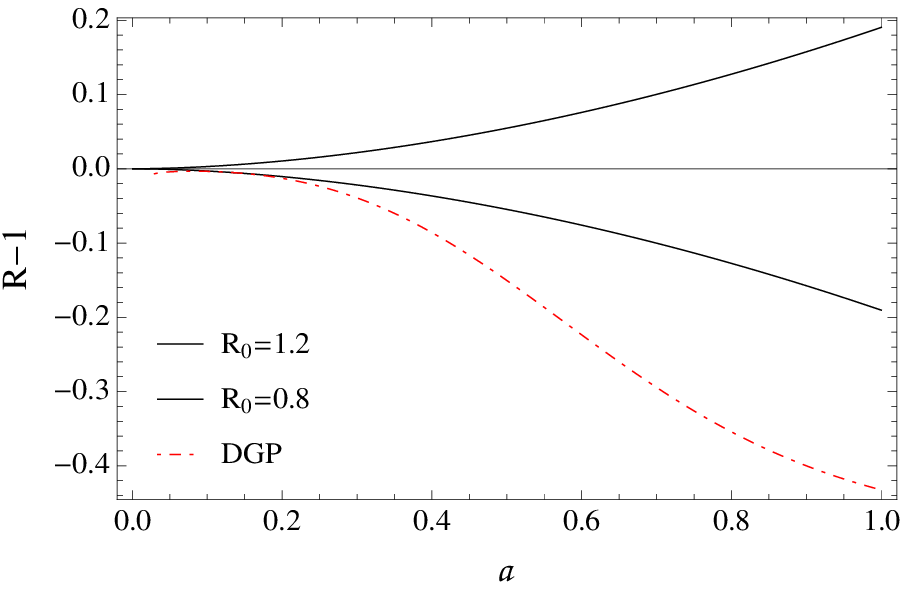}} &
{\includegraphics[width=3.45in,height=1.92in,angle=0]{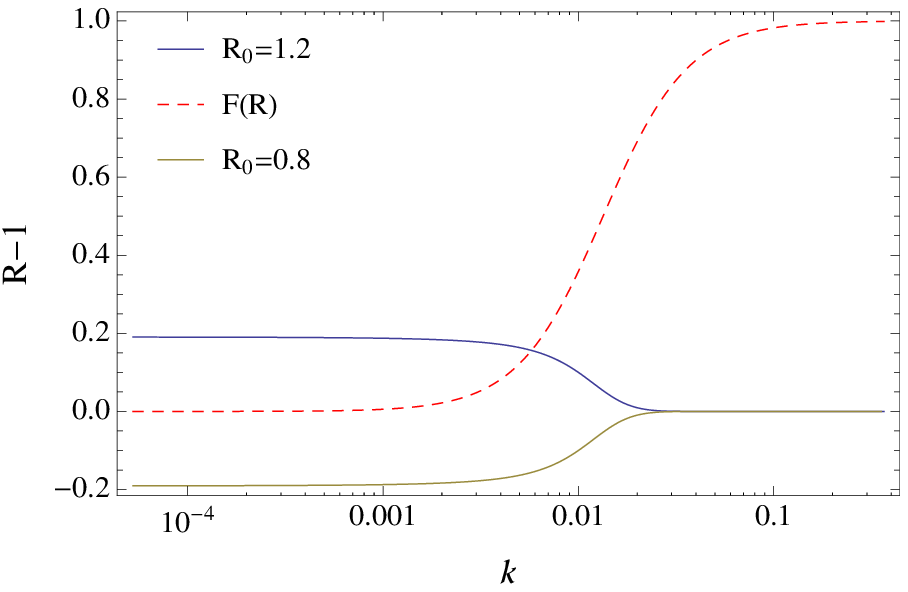}}\\ \hline
\end{tabular}
\caption{\label{f:Rplot} 
We plot the evolution of $R-1$ for Models II and III where the behavior of $R$ is given by Eq. (\ref{eq:MGR}). LEFT: We plot $R-1$ for large scales as a function of scale factor $a$. For comparison, also included in these plots is $R_{DGP} = (1+2\Omega_m(a)^2)/(2+\Omega_m(a)^2)$ for a DGP model with $\Omega_m = 0.251$ and an expansion history identical to that of $\Lambda$CDM.  One can see that DGP model deviates much more significantly than the two models plotted, thus dark energy models with anisotropic stress perturbations that could produce an MG parameter $R$ of the given amplitude would still be distinguishable from modified gravity models such as the DGP model.  RIGHT: We plot $R-1$ as a function of wave-number $k$ today.  Also included is a plot of $R-1$ for a $f(R)$ model as described by Eq. (\ref{eq:RFR}). Not only is the deviation that manifests for $f(R)$ more significant than the deviations for dark energy models with perturbations, but $f(R)$ also shows deviations mostly in a different range of $k$ values.} 
\end{figure}

In Fig. \ref{f:Rplot}, we plot $R-1$ for Models II and III. The behavior of $R$ for these models has of course already been given by Eq. (\ref{eq:MGR}). For comparison, we also include a plot of $R_{DGP}$, which is given by Eq. (\ref{eq:RDGP}) and $R_{f(R)}$ as given by Eq. (\ref{eq:RFR}).  Once again the behavior of $R$ for these two models do not deviate from $1$ as significantly as the DGP model or the $f(R)$ model.  Also for the $f(R)$ models the deviations in $R$ at mostly different scales compared to the dark energy models. Thus these dark energy models would be distinguishable from such modified gravity models. 

\subsection{Effects of changing the sound speed of dark energy perturbations\label{sec:cs2}}
In our final analysis section of the paper we would like to quickly explore the effect of the effective sound speed of dark energy perturbations, $c_s^2$, on the various tests we have discussed.  For brevity we only look at the two most extreme cases of the dark energy equations of state we have considered those with $w = -0.65$ and $w=-1.35$.  We consider a range of sound speeds, $c_s^2 = 1$, $c_s^2 = 0.1$, $c_s^2 = 0.01$, and $c_s^2 = 0$.

\subsubsection{Impact on the growth index \label{sec:cs2mggi}}
\begin{figure}[t!]
\centering
\begin{tabular}{|c|c|}
\hline
{\includegraphics[width=3.45in,height=1.92in,angle=0]{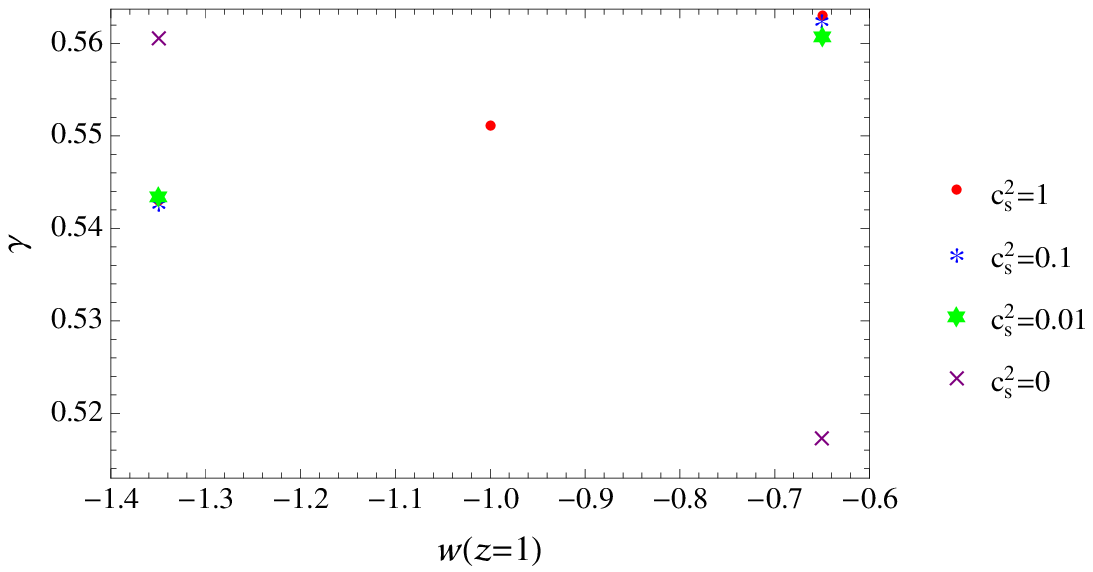}} &
{\includegraphics[width=3.45in,height=1.92in,angle=0]{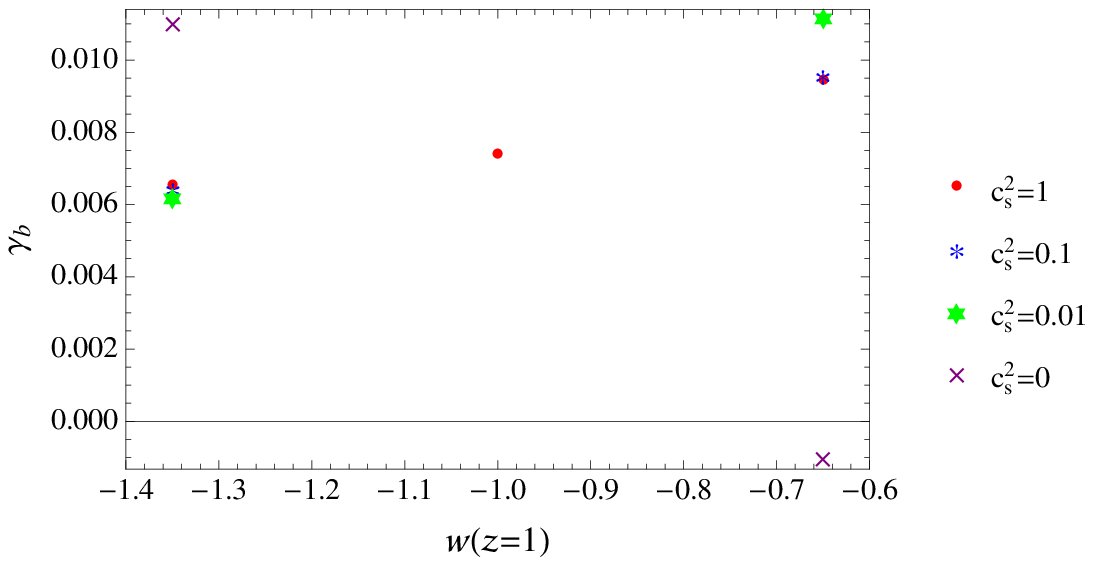}}\\ \hline
\end{tabular}
\caption{\label{f:cs2gamma} 
We plot the best fit values for the growth index parameters as versus $w$ evaluated at $z=1$ where we allow for dark energy perturbations with various values of $c_s^2$.  LEFT: We fit a constant $\gamma$ to our obtained logarithmic growth rate $f$ via the usual ansatz $f(z)=\Omega_m(z)^\gamma$.   Notice for models with $c_s^2 = 0$ the trend is reversed compared to other models with other values of $c_s^2$. RIGHT: We plot the best fits for the parameter $\gamma_b$ from the parameterization for $\gamma$ given by Eq. (\ref{eq:gammaexp}).  For the model with $c_s^2 = 0$ the value of this parameter does go negative for $w = -0.65$.  This unfortunately shows that the sign of the parameter $\gamma_b$ is not a feature that can consistently be used to distinguish between dark energy and modified gravity models.} 
\end{figure}

As we have done previously we will first explore the impact of the various values of $c_s^2$ on the value of the growth index parameter. We do not plot the behavior of the growth index parameters for dark energy models with anisotropic stress perturbations for the various $c_s^2$ considered because the changed sound speed was found to have no effect compared to the values obtained in the previous section.  However, for dark energy models that have no anisotropic stress perturbations the values of $\gamma$ were affected.  We plot these results in Fig.\ref{f:cs2gamma}.  As one can see, for most of the values of $c_s^2$, the values of the growth index parameters are clustered very close to the $c_s^2 =1$ case.  The only exception is $c_s^2 = 0$. In this case the values of both $\gamma$ and $\gamma_b$ are shifted towards higher values for $w<-1$ and towards lower values for $w>-1$.  

Once again, though, the values of $\gamma$ exhibited for all models are not very far deviated from the theoretical  value of $6/11$ and is certainly not near the values exhibited by the various modified gravity models we have discussed previously in this paper.  This leads us to a major conclusion: the constant growth index $\gamma$ is a very robust way to distinguish between dark energy models -- even extreme dark energy models --  and modified gravity models. 

The same level of robustness unfortunately does not exist for the growth index slope parameter $\gamma_b$.  Previously it was shown that the sign of this parameter could be used to distinguish between dark energy and modifications to gravity.  However, we have seen here that some extreme models of dark energy can have a sign for this parameter that is opposite to what was previously expected for dark energy models.

\subsubsection{Impact on the MG parameter $Q$ \label{sec:cs2mgq}}
\begin{figure}[t!]
\centering
\begin{tabular}{|c|c|}
\hline 
{\includegraphics[width=3.45in,height=1.92in,angle=0]{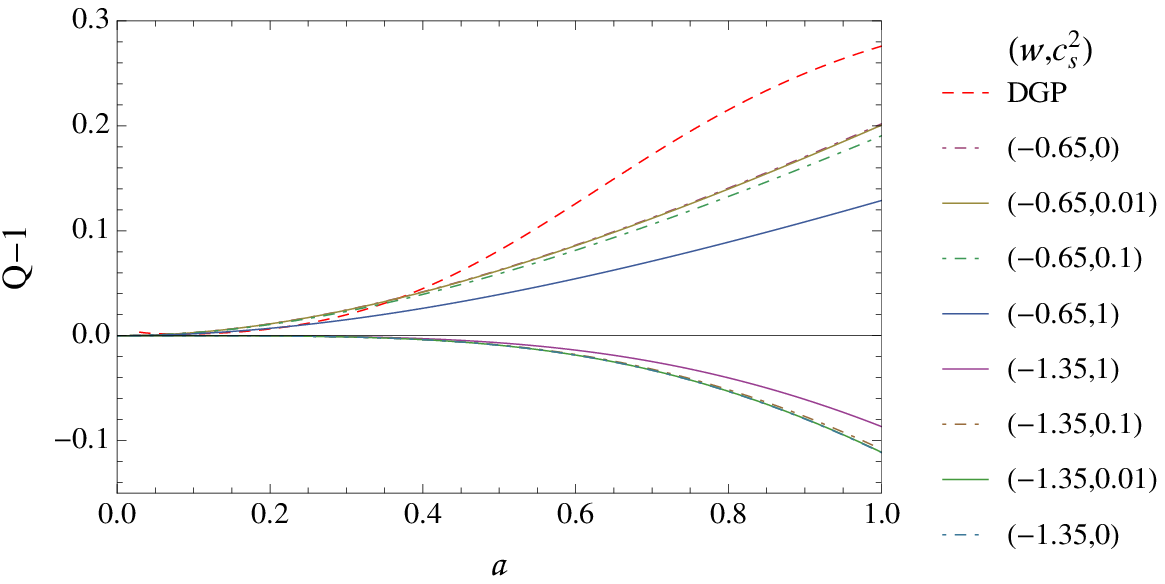}} &
{\includegraphics[width=3.45in,height=1.92in,angle=0]{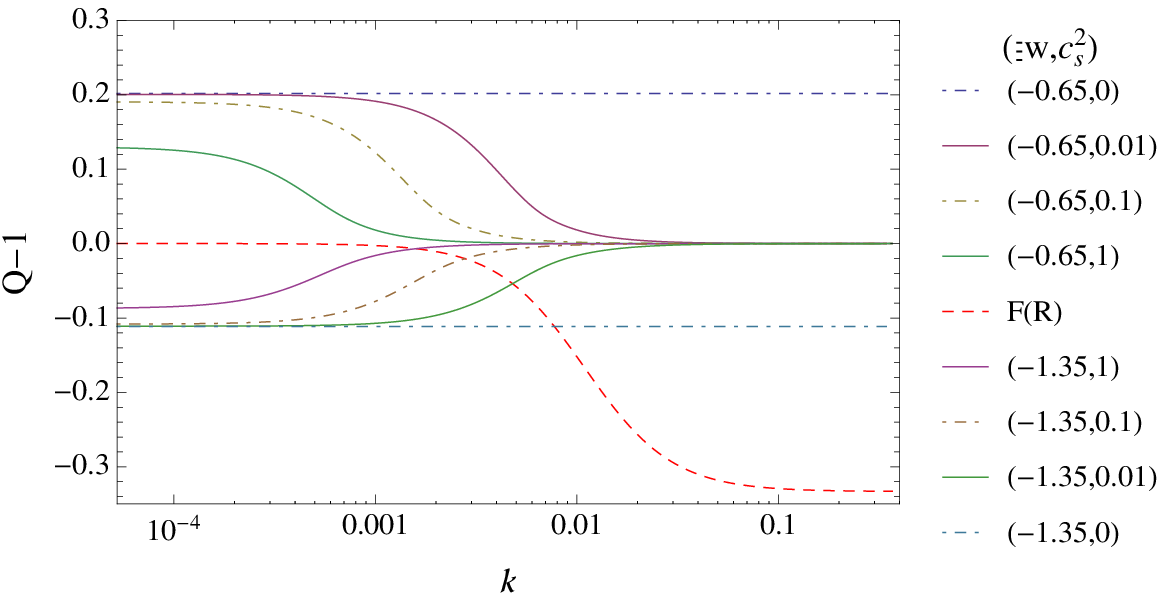}}\\ \hline
\end{tabular}
\caption{\label{f:cs2Q} 
We plot $Q-1$ as evaluated from Eq. (\ref{eq:Qeval}) for various dark energy models where we allow for dark energy perturbations with various values of $c_s^2$. The legend lists the various models used and is ordered according to their values today.  LEFT: Here we plot $Q-1$ as a function of scale factor $a$ for the scale corresponding to 40 times that of the horizon as explained in subsection (IV.A.2).  We also include a plot of $Q-1$ for a DGP model with $\Omega_m = 0.251$ and an expansion history identical to that of $\Lambda$CDM, as given by Eq. (\ref{eq:QDGP}).  This model deviates much more significantly than do any of the dark energy models with perturbations.  RIGHT: We plot $Q-1$ as a function of wave-number $k$ today.  Also included is a plot of $Q-1$ for a $f(R)$ model as described by Eq (\ref{eq:QFR}). The deviation that manifests for $f(R)$ more significant than the deviations for dark energy models with perturbations, and in most cases $f(R)$ also shows deviations for a different range of $k$ values. } 
\end{figure}

The impact of the various values of the sound speed of dark energy perturbations on the MG parameter $Q$ is quite a bit more elaborate than its impact on the growth index.  The results for dark energy models without anisotropic stress are shown in Fig.\ref{f:cs2Q}.  One quickly notices two things.  First, the absolute value of $Q$ for the models with a smaller $c_s^2$ are larger.   Second and more significantly, $Q$ deviates for a larger scale range as $c_s^2$ gets smaller.  In fact, for $c_s^2 = 0$, the value of $Q$ becomes scale independent (As an aside, this scale independence seen in $Q$ helps explain the behavior seen in the growth index as now the growth at the scales we are evaluating the growth index have obviously been modified).   At first look, this is discouraging since we can no longer say that deviations in $Q$ do occur at distinct scales from those for $f(R)$.  However we must consider the overall scale dependence and magnitude of the deviations.  These features are still distinct as the dark energy models still do not show deviations as significant as those seen in the modified gravity models shown.

\begin{figure}[t!]
\centering
\begin{tabular}{|c|c|}
\hline 
{\includegraphics[width=3.45in,height=1.92in,angle=0]{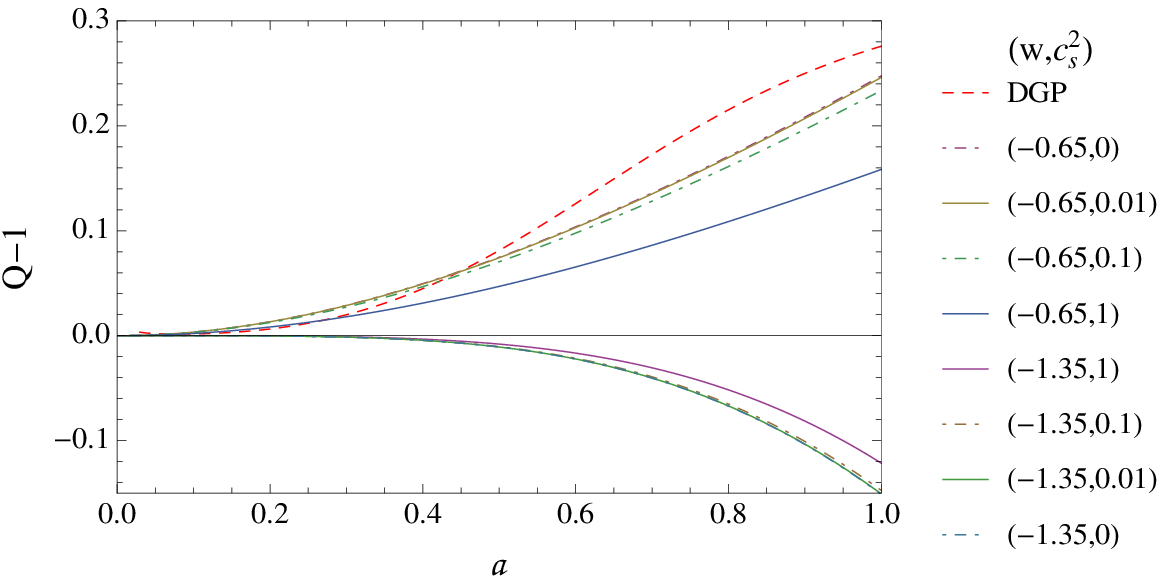}} &
{\includegraphics[width=3.45in,height=1.92in,angle=0]{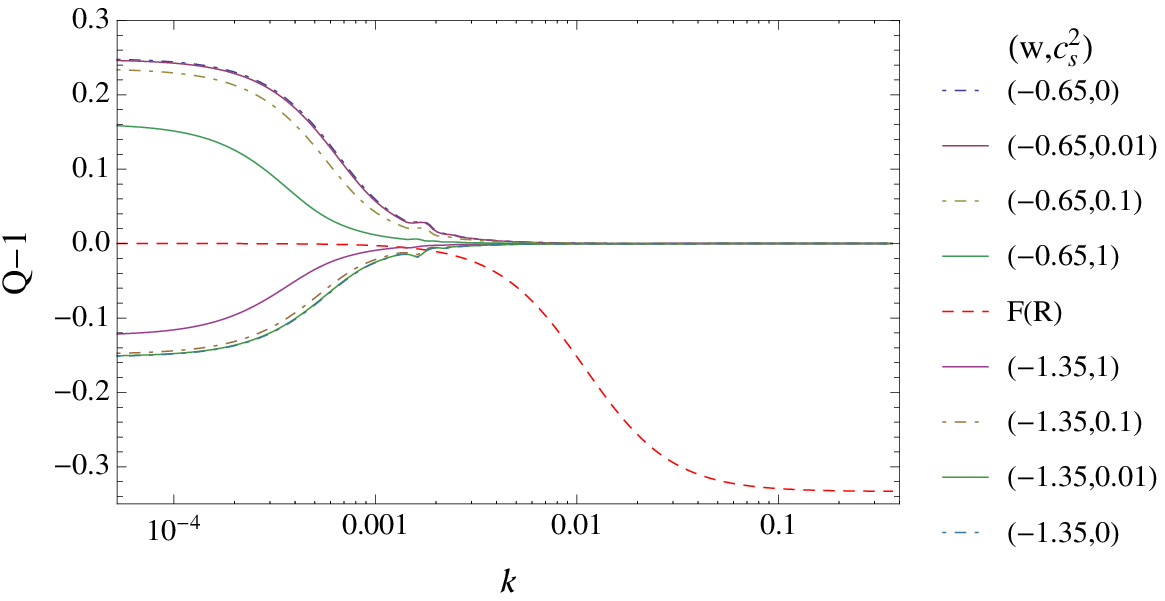}}\\ \hline
\end{tabular}
\caption{\label{f:cs2piQ} 
We plot $Q-1$ as evaluated from Eq. (\ref{eq:Qeval}) for various dark energy models where we allow for dark energy perturbations with various values of $c_s^2$ and also include anisotropic stress perturbations as described by Model I. The legend lists the various models used and is ordered according to their values today. Contrary to the behavior seen in Fig.\ref{f:cs2Q}, $Q-1$ does not become scale independent for $c_s^2=0$. LEFT: Here we plot $Q-1$ as a function of scale factor $a$ for the scale corresponding to 40 times that of the horizon as explained in subsection (IV.A.2).  We also include a plot of $Q-1$ for a DGP model with $\Omega_m = 0.251$ and an expansion history identical to that of $\Lambda$CDM, as given by Eq. (\ref{eq:QDGP}).  This model deviates much more significantly than do any of the dark energy models with perturbations.  RIGHT: We plot $Q-1$ as a function of wave-number $k$ today.  Also included is a plot of $Q-1$ for a $f(R)$ model as described by Eq (\ref{eq:QFR}). Not only is the deviation that manifests for $f(R)$ more significant than the deviations for dark energy models with perturbations, but $f(R)$ also shows deviations for a different range of $k$ values.} 
\end{figure}

In Fig.\ref{f:cs2piQ} we plot the effect of the various values of $c_s^2$ on $Q$ for models of dark energy that  have anisotropic stress perturbations described by model I.  Interestingly, the addition of anisotropic stress perturbations removes the varying scale dependence seen for the dark energy models that did not have anisotropic stress perturbations.  This is again due to the way $\delta_{\rm DE}$ and $\theta_{\rm DE}$ are affected by $\Pi_{_{\rm DE}}$ in conjunction with the scale ranges these variables contribute to $\Delta_{\rm DE}$ and thus $Q$.  Since $\Pi_{_{\rm DE}}$ acts to bring $\delta_{\rm DE}$ closer to zero at larger k-values (smaller scales) and $\delta_{\rm DE}$ is the dominant contribution to $\Delta_{\rm DE}$ in that scale range, the value of $Q$ is suppressed at smaller scales, thus restoring its scale dependence.  

As with the dark energy models without anisotropic stress though, the value of $Q$ for the large scales is increased for models with a lower $c_s^2$.  In fact, for the model with $w=-0.65$ $Q$ does begin to approach the value exhibited for the DGP model.  However, just as in the previous case, when we consider the magnitude and scale dependence of $Q$ (since DGP is mostly scale independent) we are once again fully able to distinguish between the dark energy model and the modified gravity models.  

We will quickly mention now that the impact of changing $c_s^2$ on $R$ is very small.  At large scales the value of $R$ is unchanged and the scale dependence is nearly identical to what was shown in section \ref{sec:PPmgr}.   

Given our observations on the impact of the various features of dark energy perturbations on the MG parameters $Q$ and $R$ in the above sections, it can be concluded that joint constraints on the MG parameters $Q$ and $R$ should still be able to distinguish between dark energy models including those with density and anisotropic stress perturbations and modifications to gravity, even for extreme cases.  This is especially true when we consider both the magnitude and scale dependence of these parameters.

\section{Conclusions\label{sec:conclude}}
In this work we have studied how more complex models of dark energy can affect tests that are used to distinguish between dark energy and modifications to gravity as causes for the observed cosmic acceleration. We considered dark energy models with density, velocity and three different models of anisotropic stresses perturbations. Our analysis did not include dark energy models where exotic interactions with other matter species were present. We particularly focused on two tests used to accomplish this task: the growth index parameter ,$\gamma$, which characterizes the logarithmic growth rate of perturbations, $f=d\ln\delta/d\ln a$; and the MG parameters $Q$ and $R$, which go directly into the growth equations from the perturbed FLRW metric.

We found that the growth index parameter is a robust test even when dark energy is allowed to have density perturbations or anisotropic stress perturbations.  That is, the dispersion in $\gamma$ for such dark energy models remains small even for the most extreme cases. A constant growth index was found to be particularly robust to the various extreme dark energy models we considered.  For the most extreme case it varies by only $5\%$  from its theoretical predicted LCDM value of $6/11$.  In most cases though, it follows a trend very close to the relation $\gamma = 0.552 + 0.025(1+w(z=1))$ which is consistent with the relation found in \cite{Linder2005DEP} where such perturbations were not considered.  

We also found that the sign of the rate of change of a redshift dependent index parameter is not quite as robust a test as the constant growth index.  While the parameter we used to quantify the redshift dependence in this paper maintains positive values most of the dark energy models we considered, some of the extreme models did display negative values.   Thus the approach of using the sign of this slope parameter as a test of dark energy cannot be said to be a completely reliable approach to distinguish between dark energy and modified gravity.  However, since the negative values for this parameter are only found for very extreme dark energy models that have equations of state ruled out at at least the $2\sigma$ level by current cosmological observations, one could still use the sign of this parameter as an indication that a modified gravity model might need to be considered. 

In our exploration of the impact of dark energy perturbations on the MG parameters $Q$ and $R$, we derived analytic expressions that relate these parameters to dark energy perturbation quantities.  Using the derived expressions, we looked at the effect various dark energy models would have on $Q$ and found that, while dark energy perturbations do cause $Q$ to deviate from its GR unity value, the magnitude of these deviations does not approach those exhibited by various modified gravity models such as the DGP model or the $f(R)$ gravity models. Additionally, in comparison to the $f(R)$ gravity models, the deviations for dark energy models with a non-zero effective sound speed are found to manifest at distinct scales, i.e. distinct ranges of wave-number. When exploring the impact of the dark energy perturbations on the $R$, which is only affected if dark energy has anisotropic stress perturbations, we found, just as with $Q$, that dark energy models with perturbations explored did not cause $R$ to deviate as significantly as the DGP or $f(R)$ modified gravity models.  Again, the $f(R)$ modified gravity model not only exhibited much more significant for this parameter deviations than the dark energy models, the deviations also occurred at completely distinct scales. So, additionally, the scale dependence of the MG parameters in some modified gravity model provide a solid discriminant to distinguish these models from dark energy models regardless of any perturbations.    

The findings of this paper indicate that the two tests discussed above that are used to distinguish between dark energy models and modifications to gravity are robust to dark energy density and anisotropic stress perturbations even for most extreme cases of dark energy models. Among these parameters, the growth index is found to be the most robust to these perturbations. 

\begin{acknowledgments}
We would like to thank D. Parkinson for useful discussions.  M.I. acknowledges that this material is based upon work supported by the Department of Energy (DOE) under grant DE-FG02-10ER41310, NASA under grant NNX09AJ55G, and that part of the calculations for this work have been performed on the Cosmology Computer Cluster funded by the Hoblitzelle Foundation.  J.D. acknowledges that this research was supported in part by the DOE Office of Science Graduate Fellowship Program (SCGF), made possible in part by the American Recovery and Reinvestment Act of 2009, administered by ORISE-ORAU under contract no. DE-AC05-06O100.
\end{acknowledgments}


\begin{thebibliography}{}
\bibitem{LSS04DEP} A. Lue, R. Scoccimarro, and G. Starkman, \phrd \textbf{69}, 124015 (2004), arXiv:astro-ph/0401515.
%
\bibitem{Song2005aDEP} Y. S. Song, \phrd \textbf{71}, 024026 (2005), arXiv:astro-ph/0407489.
%
\bibitem{Linder2005DEP}E. Linder, \phrd \textbf{72}, 043529 (2005), arXiv:astro-ph/0507263.%
%
\bibitem{Ishak2006DEP} M. Ishak, A. Upadhye, and D. Spergel, \phrd \textbf{74}, 043513 (2006), arXiv:astro-ph/0507184.
%
\bibitem{KST06DEP} L. Knox, Y.S. Song, and J. A. Tyson, \phrd \textbf{74}, 023512 (2006), arXiv:astro-ph/0503644.
%
\bibitem{KoyamaDEP} K. Koyama, \jcap \textbf{03} (2006) 017, arXiv:astro-ph/0601220.
%
\bibitem{Zhang2006DEP} P. Zhang, \phrd \textbf{73}, 123504 (2006), arXiv:astro-ph/0511218.
%
\bibitem{HS07DEP}W. Hu and I. Sawicki, \phrd \textbf{76}, 104043 (2007), arXiv:0708.1190.  
%
\bibitem{KS08DEP}  M. Kunz and D. Sapone, Phys. Rev. Lett. \textbf{98}, 121301 (2007), arXiv:astro-ph/0612452.
%
\bibitem{GabadadzeDEP} G. Gabadadze and A. Igelsias, Classical Quantum Gravity \textbf{25}, 154008 (2008), arXiv:0712.4086.
%
\bibitem{WeiDEP} H. Wei and N. Zhang, \phrd \textbf{78}, 023011 (2008), arXiv:0803.3292.
%
\bibitem{HL07DEP} D. Huterer and E. Linder, \phrd \textbf{75}, 023519 (2007), arXiv:astro-ph/0608681.
%
\bibitem{LC07DEP}E. Linder and R. Cahn, Astropart. Phys. \textbf{28}, 481 (2007), arXiv:astro-ph/0701317.
%
\bibitem{Gong2008DEP} Y.G. Gong, \phrd \textbf{78}, 123010 (2008).
%
\bibitem{PolarskiDEP} D. Polarski and R. Gannouji, Phys. Lett. B. \textbf{660}, 439 (2008), arXiv:0710.1510.
%
\bibitem{MHH1DEP} M. J. Mortonson, W. Hu and D. Huterer, \phrd \textbf{79}, 023004 (2009), arXiv:0810.1744.
%
\bibitem{Dent2009DEP} J. Dent, S. Dutta, and L. Perivolaropoulos, \phrd \textbf{80}, 023514 (2009), arXiv:0903.5296.
%
\bibitem{GIW09DEP} Y. Gong, M. Ishak, and A. Wang, \phrd \textbf{80}, 023002 (2009), arXiv:0903.0001.
%
\bibitem{GMPDEP}R. Gannouji, B. Moraes, and D. Polarski, \jcap \textbf{02} (2009) 034, arXiv:0809.3374.
%
\bibitem{Ishak2009DEP} M. Ishak and J. Dossett, \phrd \textbf{80}, 043004 (2009), arXiv:0905.2470.
%
\bibitem{Simpson09DEP}F. Simpson and J.A. Peacock, \phrd \textbf{81}, 043512 (2010), arXiv:0910.3834.
%
\bibitem{WuDEP} P. Wu, H. Yu, and X. Fu, \jcap \textbf{06} (2009) 019, arXiv:0905.3444.
%
\bibitem{MotohashiDEP}H. Motohashi, A.A. Starobinsky, and J. Yokoyama, (2010), arXiv:1002.0462.
%
\bibitem{ZhangEtal2007DEP} P.J. Zhang, M. Liguori, R. Bean, and S. Dodelson, Phys. Rev. Lett. \textbf{99}, 141302 (2007), arXiv:0704.1932.
%
\bibitem{JainDEP} B. Jain and P. Zhang, \phrd \textbf{78}, 063503 (2008), arXiv:0709.2375.
%
\bibitem{BZ08DEP} E. Bertschinger and P. Zukin, \phrd \textbf{78}, 024015 (2008), arXiv:0801.2431.
%
\bibitem{Linder11DEP} E. Linder, Phil. Trans. Roy. Soc. A \textbf{369}, 4985 (2011), arXiv:1103.0282.
%
\bibitem{BakerDEP}T. Baker, \phrd \textbf{85}, 044020 (2012), arXiv:1111.3947.
%
\bibitem{AcquavivaDEP}V. Acquaviva, A. Hajian, D. Spergel, and S. Das, \phrd \textbf{78}, 043514 (2008), arXiv:0803.2236.
%
\bibitem{DossettDEP} J. Dossett, M. Ishak, J. Moldenhauer, Y. Gong, and A. Wang, \jcap \textbf{04} (2010) 022, arXiv:1004.3086.
%
\bibitem{MHH2DEP} M. J. Mortonson, W. Hu and D. Huterer, \phrd \textbf{81}, 063007 (2010), arXiv:0912.3816.
%
\bibitem{Rapetti2012DEP}D. Rapetti {\it et~al.}, \mnras \textbf{432}, 973 (2013), arXiv:1205.4679.
%
\bibitem{CCM07DEP} R. Caldwell, A. Cooray, and A. Melchiorri, \phrd \textbf{76}, 023507 (2007), arXiv:astro-ph/0703375.
%
\bibitem{SerraDEP}  P. Serra, A. Cooray, S.  Daniel, R. Caldwell, and A. Melchiorri, \phrd \textbf{79}, 101301 (2009), arXiv:0901.0917.
%
\bibitem{ThomasDEP}S. Thomas, F. Abdalla and J. Weller, \mnras \textbf{395}, 197 (2009), arXiv:0810.4863.
%
\bibitem{GMSM09}T. Giannantonio, M. Martinelli, A. Silvestri and A. Melchiorri, \jcap \textbf{04} (2010) 030, arXiv:0909.2045.
%
\bibitem{Daniel2008DEP} S. Daniel, R. Caldwell, A. Cooray, and  A. Melchiorri \phrd \textbf{77}, 103513 (2008), arXiv:0802.1068.
%
\bibitem{Daniel2009DEP}S. Daniel, R. Caldwell, A. Cooray, P. Serra, and  A. Melchiorri, \phrd \textbf{80}, 023532 (2009), arXiv:0901.0919.
%
\bibitem{Daniel2010DEP} S. Daniel, E. Linder, T. Smith, R. Caldwell, A. Corray, A Leauthaud, and L. Lombriser, \phrd \textbf{80}, 123508 (2010), arXiv:1002.1962.
%
\bibitem{Bean2010DEP} R. Bean and M. Tangmatitham, \phrd \textbf{81}, 083534 (2010), arXiv:1002.4197.
%
\bibitem{DL2010DEP} S. Daniel and E. Linder \phrd \textbf{82}, 103523 (2010), arXiv:1008.0397.
%
\bibitem{GongBo2010DEP} G. Zhao et. al. \phrd \textbf{81}, 103510 (2010), arXiv:1003.0001.
%
\bibitem{Dossett2011DEP} J. Dossett, J. Moldenhauer, and M. Ishak, \phrd \textbf{84},  023012, (2011), arXiv:1103.1195.
%
\bibitem{ISiTGRDEP}J. Dossett, M. Ishak, and J. Moldenhauer, \phrd \textbf{84}, 123001 (2011), arXiv:1109.4583.
%
\bibitem{DIDEP}J. Dossett and M. Ishak, \phrd \textbf{86}, 103008 (2012), arXiv:1205.2422.
%
\bibitem{MGCAMB2DEP} A. Hojjati, L. Pogosian, and G. Zhao, \jcap \textbf{08} (2011) 005, arXiv:1106.4543.
%
\bibitem{LombriserDEP} L. Lombriser, \phrd \textbf{83}, 063519 (2011), arXiv:1101.0594.
%
\bibitem{SongZhao2011DEP} Y. S. Song, G. B. Zhao, D. Bacon, K. Koyama, R. C. Nichol, and L. Pogosian, \phrd \textbf{84} 083523, (2011), arXiv:1011.2106.
%
\bibitem{TorenoDEP} I. Toreno, E. Semboloni, and T. Schrabback,  Astron. Astrophys. \textbf{530}, A68 (2011), arXiv:1012.5854.
%
\bibitem{Laszlo2011DEP}I. Laszlo, R. Bean, D. Kirk, and S. Bridle, (2011), arXiv:1109.4535.
%
\bibitem{HojjatiPCADEP} A. Hojjati, G. B. Zhao, L. Pogosian, A. Silvestri, R. Crittenden, and K. Koyama, \phrd \textbf{85}, 043508 (2012), arXiv:1111.3960.
%
\bibitem{Simpson2012DEP}F. Simpson {\it et~al.}, \mnras \textbf{429}, 2249 (2013), arXiv:1212.3339.
%
\bibitem{PeeblesDEP} P.J.E. Peebles, \emph{The Large Scale Structure of the Universe} (Princeton University Press, Princeton, NJ, 1980).
%
\bibitem{wangDEP}L. Wang and P. Steinhardt, Astrophys. J 508, \textbf{483}, 1998.
%
\bibitem{MaDEP}C. Ma and E. Bertschinger, Astrophys. J. \textbf{455}, 7 (1995).
%
\bibitem{ZSBDEP}M. Zaldarriaga, U. Seljak, and E. Bertschinger,  Astrophys. J. \textbf{494}, 491 (1998), arXiv:astro-ph/9704265.
%
\bibitem{Hu98DEP}W. Hu, Astrophys. J. \textbf{506}, 485 (1998), arXiv:astro-ph/9801234.
%
\bibitem{BDDEP}R. Bean and O. Dore, \phrd \textbf{69}, 083503 (2004), arXiv:astro-ph/0307100.
%
\bibitem{dPHLDEP}R. de Putter, D. Huterer, and E. Linder, \phrd \textbf{81}, 103513 (2010), arXiv:1002.1311.
%
\bibitem{KM05DEP} T. Koivisto and D. F. Mota, \phrd \textbf{73}, 083502,(2006), arXiv:astro-ph/0512135.
%
\bibitem{MKKG07DEP}D.F. Mota, J.R Kristiansen, T. Koivisto, and N.E. Groeneboom, \mnras \textbf{382}, 793 (2007), arXiv:0708.0830.
%
\bibitem{LewisCAMBDEP} A. Lewis, A. Challinor, and A. Lasenby, Astrophys. J. \textbf{538}, 473 (2000); A. Lewis and A. Challinor, Code for Anisotropies in the Microwave Background \url{http://camb.info}.
%
\bibitem{CPLCPDEP}M. Chevallier and D. Polarski, Int. J. Mod. Phys. D \textbf{10}, 213 (2001).
%
\bibitem{CPLLDEP}E. Linder, Phys. Rev. Lett. \textbf{90}, 091301 (2003).
%
\bibitem{AZBDEP}M. Amin, P. Zukin, and E. Bertschinger, \phrd \textbf{85}, 103510 (2012), arXiv:1108.1793.
%
\bibitem{WMAP9DEP}G. Hinshaw {\it et~al.}, (2012), arXiv:1212.5226.
%
\bibitem{BMDEP}P. Brax and J. Martin, Phys. Lett. B \textbf{468}, 40 (1999), arXiv:astro-ph/9905040.
%
\bibitem{WADEP}J. Weller and A. Albrecht, \phrd \textbf{65}, 103512, (2002).
%
\bibitem{WMAP5DEP}E. Komatsu {\it et~al.}, Astrophys. J. Suppl. \textbf{180}, 330 (2009), arXiv:0803.0547.
%
\bibitem{WMAP74DEP}E. Komatsu {\it et~al.}, Astrophys. J. Suppl. Ser. \textbf{192}, 18 (2011), arXiv:1001.4538.
%
\bibitem{DGPDEP} G. Dvali, G. Gabadadaze, and M. Porrati, Phys. Lett. B. \textbf{485}, 208 (2000), arXiv:hep-th/0005016.
%
\bibitem{CuscutonDEP} N. Afshordi, D. Chung, M. Doran, G. Geshnizjani, \phrd \textbf{75}, 123509, (2007).
%

\end{thebibliography}
\end{document}